# The Solar Connection of Enhanced Heavy Ion Charge States in the Interplanetary Medium: Implications for the Flux-rope Structure of CMEs

N. GOPALSWAMY

*NASA Goddard Space Flight Center, Greenbelt, MD*

P. MÄKELÄ, S. AKIYAMA, H. XIE, S. YASHIRO

*The Catholic University of America, Washington, DC*

A. A. REINARD

*NOAA Space Weather Prediction Center, Boulder, CO*

Abstract

We investigated a set of 54 interplanetary coronal mass ejection (ICME) events whose solar sources are very close to the disk center (within ±15° from the central meridian). The ICMEs consisted of 23 magnetic cloud (MC) events and 31 non-MC events. Our analyses suggest that the MC and non-MC ICMEs have more or less the same eruption characteristics at the Sun in terms of soft X-ray flares and CMEs. Both types have significant enhancements in charge states, although the non-MC structures have slightly lower levels of enhancement. The overall duration of charge state enhancement is also considerably smaller than that than that in MCs as derived from solar wind plasma and magnetic signatures. We find very good correlation between the Fe and O charge state measurements and the flare properties such as soft X-ray flare intensity and flare temperature for both MCs and non-MCs. These observations suggest that both MC and non-MC ICMEs are likely to have a flux-rope structure and the unfavorable observational geometry may be responsible for the appearance of non-MC structures at 1 AU. We do not find any evidence for active region expansion resulting in ICMEs lacking a flux rope structure because the mechanism of producing high charge states and the flux rope structure at the Sun is the same for MC and non-MC events.

## 1. Introduction

The occurrence of high charge states of elements such as oxygen, silicon, and iron at times of low solar wind kinetic temperature was attributed to heated flare plasma long ago (Bame *et al.*, 1979). The low solar wind kinetic temperature is one of the indicators of coronal mass ejections (CMEs) in the interplanetary space (i.e., ICMEs). Bame *et al.* (1979) also suggested that "magnetic bottles" might carry the flare-heated plasma with the higher charge-state ions created due to the higher temperature of the flare plasma low in the corona. Furthermore, they compared synthetic ion spectra with that of the observations and estimated a source temperature of 3.4 MK for O ions and 2.9 MK for Fe ions. The charge states are unchanged when the plasma containing heavy elements (solar wind or CME) leave the corona because the recombination time scale far exceeds the expansion time scale of the plasma. This is known as the freezing-in concept (Hundhausen, Gilbert, and Bame, 1968). Thus the charge states of heavy elements observed in the interplanetary medium preserve the coronal conditions at which they originated. Henke *et al.* (1998; 2001) suggested that the ICMEs with enhanced charge state have the magnetic cloud (MC) structure, which is the same as the flux rope. In this paper we use MC and flux rope interchangeably, but observationally, MCs are characterized by enhanced magnetic field with a smooth rotation of one of the components transverse to the Sun-Earth direction, and low values of proton temperature or plasma beta (Burlaga *et al.* 1981). Henke *et al.* (1998) analyzed 56 ICMEs observed by the *Ulysses* spacecraft and found that those with MC structure have an increased O7+/O6+ ratio (herein after referred to as O7O6) with respect to the ambient solar wind whereas non-MC ICMEs seldom show such enhancement. Furthermore, the events with enhanced O7O6 also showed an enhancement in the Fe12+/Fe11+ charge state ratio. Aguilar-Rodriguez, Blanco-Cano, and Gopalswamy (2006) considered a much larger sample of ICMEs (28 MCs and 117 non-MCs) observed at Sun-Earth L1 by the ACE spacecraft and confirmed the result of Henke *et al.* (1998; 2001). Reinard (2008) examined the source location and flare size at the Sun and the in-situ density and temperature for a large numbers of ICMEs and found that ICMEs may have a basic structure consisting of a core (or cores) of magnetic-cloud plasma surrounded by an envelope with weaker charge-state signatures. These studies indicate that the presence of enhanced charge states observed in interplanetary space is likely due to a CME at the Sun that is magnetically connected to a flare. In light of these findings, we are left to question why some ICMEs exhibit a flux rope structure while others do not.

How do we distinguish between MC and non-MC ICMEs? The simplest classification is to lump all the ICMEs that do not have flux rope structure as non-MC ICMEs. These are also referred to as non-cloud ICMEs or ejecta. The flux rope is thought to be formed out of reconnection during the eruption process and is observed as an MC in the interplanetary medium (see *e.g.*, Qiu *et al.* 2007). On the other hand, it is possible that a set of loops from an active region on the Sun can simply expand into the IP medium and can be detected as an enhancement in the magnetic field with respect to the ambient medium (Gosling, 1990) without any flux rope structure. Clearly, the magnetic signatures will be different in the two cases. A spacecraft passing through the flux rope will see a smooth rotation of the magnetic field throughout the body of the ICME, while the expanded loop system will show no rotation. If we take just the IP observations, we may be able to explain MCs as flux ropes and non-MCs as expanding loops. However, they should show different charge state characteristics because of the different solar origins. The flux rope forms during the flare process and hence is accessed by the hot plasma resulting in high charge states inside MCs when observed at 1 AU. Expanding loops on the other hand should not have high charge states because there may not be any reconnection involved (Uchida *et al.* 1992). Under such a scheme, the non-MC events should not have a flare association and the associated CME, if any, is expected to be generally slow. However, all the non-MC ICMEs are also associated with flares and the corresponding white-light CMEs are fast and wide (Gopalswamy *et al.*, 2010a,b).

An alternative approach is to understand the difference between MCs and non-MCs as a direct consequence of the observing geometry. According to this view, all ICMEs are flux ropes, but they do not appear so if they are not heading towards the observer (Marubashi, 1997; Owens *et al.*, 2005; Gopalswamy, 2006a; Riley *et al.*, 2006). Gopalswamy (2006a) and Gopalswamy *et al.* (2009a) compared the solar source locations of MCs, non-MCs, and shocks not followed by discernible ejecta ("driverless" shocks) and found a distinct pattern. As one moves from the disk center to the limb, one first encounters mostly MCs, then mostly non-MC ICMEs, and finally the driverless shocks. MCs are associated with CMEs heading directly towards Earth. The shocks without discernible ejecta are due to CMEs ejected almost orthogonal to the Sun-Earth line. This gives a clue that the CMEs ejected at intermediate angles may turn up as a non-MCs for an observer along the Sun-Earth line. So, viewing angle may be the reason that certain ICMEs do not have a flux-rope structure. Gopalswamy *et al.* (2009a) noted two major exceptions to this pattern. (i) There are some driverless shocks from the disk center. This was shown to be due to the deflection of CMEs by nearby coronal holes. (ii) There are too many non-MC ICMEs that have their solar sources close to the disk center, contradicting the geometrical approach. In this paper, we examine these disk-center events in more detail to see if the geometrical approach still holds and why they deviate from the geometrical hypothesis.

Two Coordinated Data Analysis Workshops (CDAWs) addressed this central question: Do all ICMEs contain a flux rope structure? Solar and interplanetary data from space and ground based instruments were assembled and analyzed during the CDAWs to answer this question. Data analyses were combined with modeling near the Sun as well as in the interplanetary medium to check if observing geometry is responsible for not observing the flux rope structure. In this paper, we make use of the charge state information of ICMEs to address the question of flux rope structure of CMEs.

## 2. Data Description

The CDAW events were extracted from the list of shock-driving ICMEs published in Gopalswamy *et al.* (2010a) in the electronic supplement (http://iopscience.iop.org/0004-637X/710/2/1111/fulltext/apj_710_2_1111.tables.html) with the criterion that the solar sources of the ICMEs should be within the longitude range ±15°. There are 59 events meeting this criterion, but further examination revealed that the solar sources had to be revised in 5 cases reducing the number of events to 54, of which 23 are MCs and the remaining 31 are non-MC ICMEs. According to the geometrical hypothesis, all the CMEs originating from close to the disk center should be observed as a flux rope by an Earth observer. Obviously this is not the case. We attempt to find out why using flare and CME observations near the Sun and charge state observations of ICMEs near Earth.

This paper uses two measures of charge states in analyzing MC and non-MC structures. The first one is the average Fe charge state denoted by QFe (see Lepri *et al.*, 2001) and is given by $\sum n_i Q_i$, where $n_i$ is the density of the Fe ions with charge state $Q_i$ (the subscript $i$ numbers the Fe charge states present in the plasma). The density is normalized such that $\sum n_i = 1$. As Lepri *et al.* (2001) showed, QFe ~ 11 corresponds to the slow solar wind. QFe>11 indicates hotter plasma typically found inside ICMEs (see also Lepri and Zurbuchen, 2004). The second measure of charge states is the ratio of densities of O ions ionized 7 and 6 times (O7+ and O6+), denoted by O7+/O6+ or simply O7O6 (Henke *et al.*, 1998; 2001; Aguilar-Rodriguez, Blanco-Cano, and Gopalswamy, 2006; Reinard 2005; 2008). The average value of O7O6 is ~0.3 in the slow solar wind (See Zhao, Zurbuchen, and Fisk, 2009 for the range of O7O6 values in different types of solar wind). We take twice this value (0.6) as the threshold to indicate ICME plasma. In previous papers, slightly larger values (0.7, 0.8 or 1) have been used to minimize the number of false identifications (see *e.g.*, Reinard, 2008). Here we are concerned with maximizing the number of enhancements in identified ICMEs, so 0.6 is justified.

A typical ICME event analyzed in this paper has a leading shock followed by an interval of ICME identified from plasma and magnetic (plasmag) signatures. For identifying an ICME, the primary characteristic used is the depressed solar wind proton temperature (a plasma signature). In addition, magnetic signatures such as enhanced field strength and smooth rotation of the vertical or azimuthal component are used to identify a MC event. We also refer to MC events as flux rope events. Figure 1 shows the O7O6 and QFe values for two events that occurred in quick succession, taken from the CDAW list. The sheath following the shock S1 has low charge state values, similar to the upstream plasma. At the first ICME (EJ1) boundary, the charge states climb to large values. The peak value of O7O6 in the EJ1 interval is 3.1 and the average value is 1.6. Similarly, the peak and average values of QFe are 14.2 and 12.9, respectively. All these numbers are above the threshold values set above and hence represent the hot plasma from the flare site that entered into the ICME when it formed near the Sun. The rear boundary of EJ1 is not clear because it coincides with the second shock S2 driven by the second ICME (EJ2). Both O7O6 and QFe show enhancements in the downstream of S2. According to the charge state signature, the rear boundary of EJ1 should be around 18 UT on 2000 July 11, which is only a few hours ahead of EJ2. Clearly, S2 has penetrated into EJ1 and the sheath of S2 is mostly EJ1. In this case, the sheath of S2 will have enhanced charge state, but it is not the property of the sheath; the origin is the preceding ICME. The QFe is enhanced and relatively smooth within EJ2 with peak and average values of 16.4 and 14.9, respectively. On the other hand, the O7O6 is fluctuating with at least five peaks, which seems to be a characteristic of many O7O6 events.  The duration of O7O6 is also slightly lower than that of QFe. The actual duration of O7O6 is even smaller if we exclude intervals when O7O6 drops below 0.6. The peak and average O7O6 are 2.0 and 1.0, respectively. Following this procedure, we compute the following quantities for each of the CDAW events: (i) the peak and average QFe within the ICME interval identified by plasmag signatures, (ii) the peak and average O7O6 within the ICME interval, (iii) the charge state duration ignoring the rear boundary of ICME (similar to EJ1 in Figure 1, where the charge state signatures extend beyond the EJ1 boundary obtained from plasmag signatures), and (iv) the duration within the ICME boundary when the charge state remains above the threshold. We analyze these six parameters for MC and non-MC events taken separately and as a combined set.

We also compile the properties of CMEs associated with the ICMEs as observed by the *Large Angle and Spectrometric Coronagraph* (LASCO) on board the *Solar and Heliospheric Observatory* (SOHO) and listed in the on line CME catalog (http://cdaw.gsfc.nasa.gov/CME_list, see Yashiro *et al.*, 2004; Gopalswamy *et al.* 2009b). We specifically use CME speed, apparent angular width, and acceleration without correcting for projection effects.

Finally, we compile the flare properties of the CMEs such as the flare size given by the peak soft X-ray flux ($W/m^2$) in the 1 – 8 Å GOES channel (used to classify the flare importance). Since the flare temperature is an important quantity that decides the charge heavy-ion charge state in the flare plasma that enters into the CMEs, we compute it using the method outlined by Garcia (1994). The method involves obtaining the ratios of soft X-ray flux in the 1 – 8 Å and 0.5 – 4 Å GOES channels to get the temperature. A software routine is available in the SolarSoft, which we make use of in obtaining the flare temperature.

Table 1 shows the list 59 events selected for the two CDAW sessions. Column 1 gives the original serial number of the events used in the CDAW sessions. The date and time of the interplanetary shocks are given in columns 2 and 3. Information on the shock-driving ICMEs is given in columns 4–8 with the ICME type (MC for magnetic clouds and EJ (ejecta) for non-MC ICMEs in column 4) followed by the start and end times. Information the white-light CMEs identified in the field of view of the SOHO/LASCO telescopes is given in columns 9–13 with date and time followed by CME properties (width, speed, and acceleration).

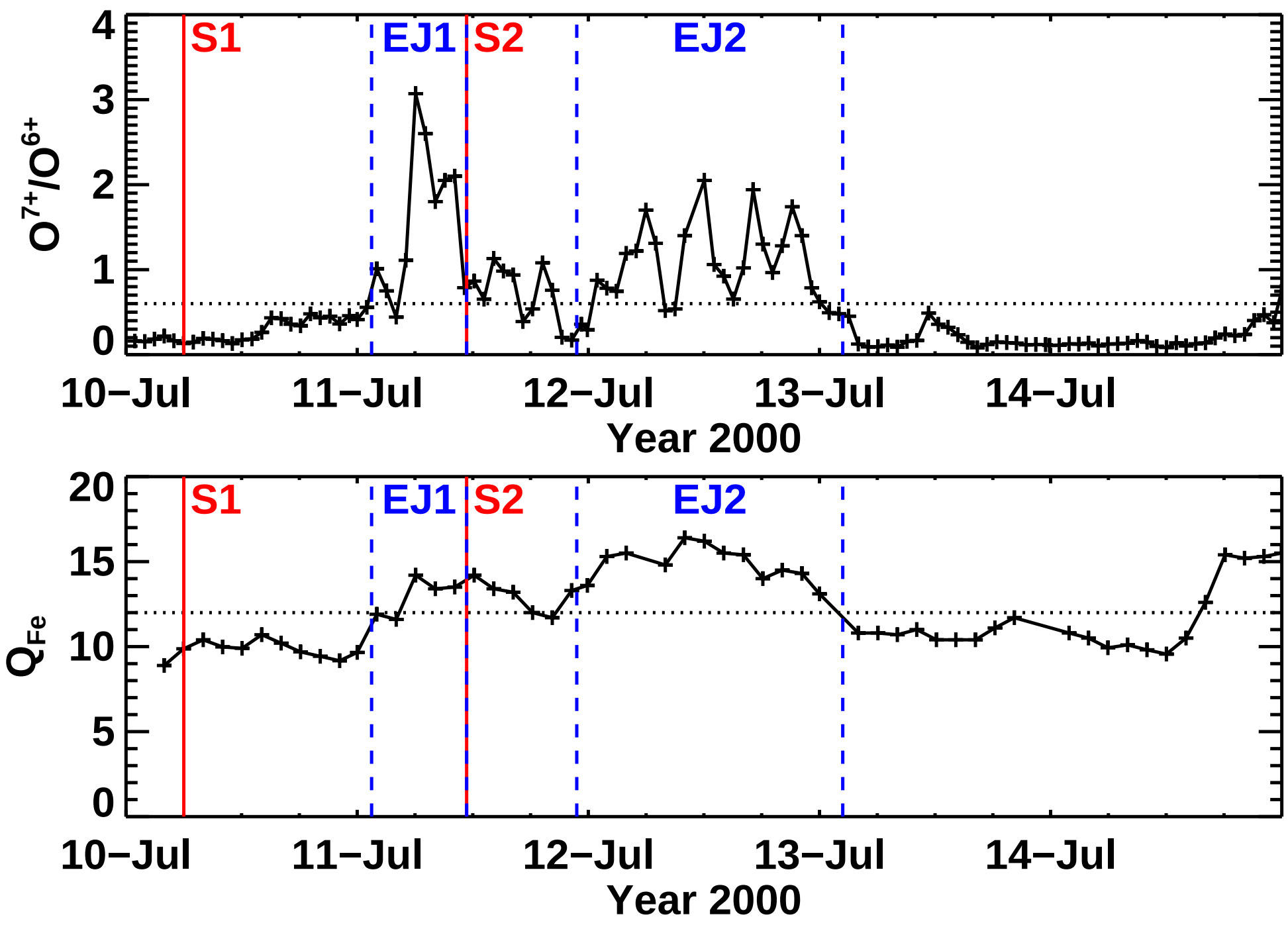

Figure 1. Charge-state time profile of the 2000 July 10 and 11 ICME events with O7O6 ratio (top) and QFe (bottom) plotted with a 1-h and 2-h time resolution, respectively . The boundaries derived from plasma and magnetic signatures of the ICMEs (EJ1, EJ2) are denoted by the vertical dashed lines. EJ1 happens to be second largest O7O6 event among the non-MC events. The leading shocks (S1, S2) of the ICMEs are denoted by the vertical solid lines. Clearly the two ICMEs are very close to each other, with the second shock already inside the first ICME. In fact, the sheath of the second shock consists mostly of the first CME.

Columns 14–16 give the solar source information of the CMEs: flare onset, flare location (heliographic coordinates), and the soft X-ray flare importance. If the associated flare is not seen above the background, the onset time of the associated eruptive prominence (EP) or post-eruption arcade (PEA) is listed with EP or PEA entered in the flare importance column. Column 17 indicates whether the event is associated with type II bursts in the metric and/or longer wavelength domains. Columns 18 – 23 give the Fe charge state information: QFe peak, QFe averaged over the event duration, duration of QFe enhancement from the first plasmag boundary until the charge state drops to the background level (dur1), cumulative duration of QFe enhancement above the threshold value of 12 (dur2), ratio of dur1 to the plasmag duration of the ICME, and the ratio of dur2 to the plasmag duration. Columns 24-29 give the same information as in columns 18-23, but for O7O6. We analyze these data to understand the difference between MC and EJ-associated CMEs and how the results can be used to find out if all CMEs have a flux rope structure.

## 3. Analysis and Results

Several results can be directly extracted from Table 1. (i) Out of the 23 MC events, two had QFe data gaps. Of the remaining 21 events, 20 had peak QFe ≥ 12.0. In all these cases, there was a definite increase in QFe sometime during the MC interval obtained from plasma signatures. Only one event did not have any QFe signature (the QFe value remained the same before the shock, in the sheath, and in the MC interval). This means, 95% of the MC events had QFe enhancement. Three of the 31 non-MC events had QFe data gap. Out of the remaining 28, only six events had QFe < 12.0, which means 79% of the non-MC events had QFe enhancement. If we use the nominal solar wind value of QFe=11, then only three non-MC events had QFe<11, indicating ~89% of non-MC events having high charge state. This is only slightly smaller than what was found in the MC events. (ii) The O7O6 within the ICME interval exceeded 0.6 in all but one of the MC events, which means 95% of the MC events had enhanced O charge state ratio. On the other hand, eight of the non-MC events had O7O6 ratio <0.6, which means about 73% of the EJ events had enhanced O7O6 during the ICME interval. These two results suggest that most of the non-MC events behave similar to the MC events in terms of the enhanced QFe and O7O6 during the ICME interval. (iii) All but three of the non-MC events have a '+' sign following the 'EJ' symbols in column 3 of Table 1. EJ+ means it was possible to fit a flux rope to the solar wind data of these ICMEs by adjusting the boundary of the ICMEs and using either a cylindrical or toroidal geometry for the flux rope (see Marubashi *et al.* 2012, under preparation, for more details on the flux rope fitting). This result is consistent with the fact that most of the ICMEs have QFe and O7O6 ratio increases within the ICME interval. Of the three "EJ-" events, two were associated with weak flare signatures and no charge state enhancement, and the third had marginal charge state enhancement. These three events are discussed in more detail in section 3.5.

### 3.1 Charge State Distributions

Figure 2 shows the QFe distributions inside all ICMEs in the CDAW list in comparison with MC and non-MC events. The mean (13.2) and median (13.5) QFe values of the combined set clearly exceed the nominal slow solar wind value (11). The corresponding values for MC and non-MC events lie above and below those of the combined set. Note also that all the mean and median values are at or above the nominal solar wind values. In the distribution of average QFe, the lower mean value results because there are intervals of low charge state during the ICME interval, when QFe dropped below the threshold value. In addition, we see that highest QFe was attained in MCs, but only in the next bin (17.5 *vs*. 16.5 for peak QFe and 16.5 *vs*. 15.5 for average QFe).

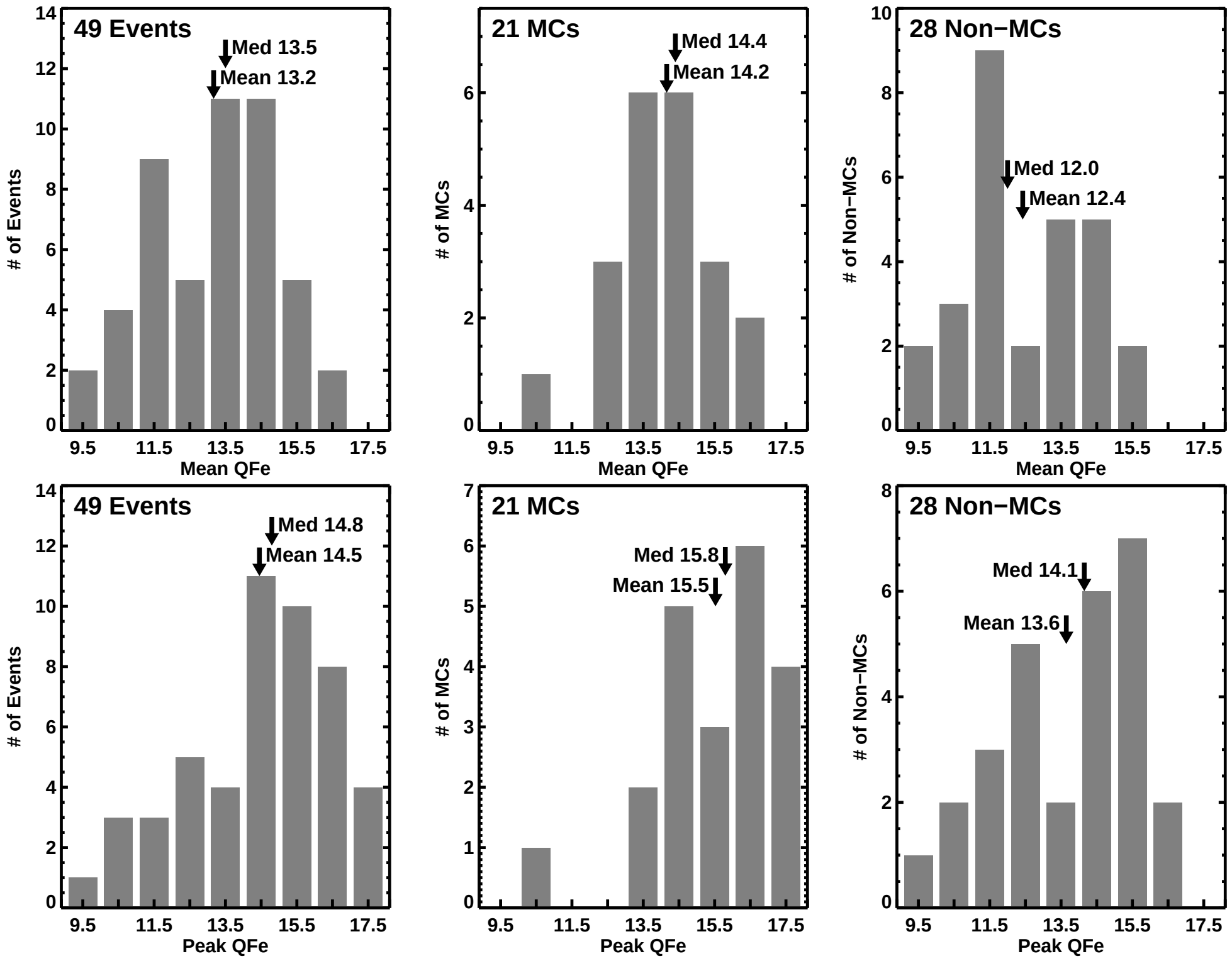

Figure 2. Distribution of QFe inside ICMEs using average (top) and peak (bottom) values within the ICME intervals. MCs and non-MC ICMEs distinguished. The mean and median values are shown on the plots.

The distributions of O7O6 values follow a pattern similar to the QFe values. MCs clearly have the highest O7O6. When peak O7O6 inside the ICME intervals are considered, MC intervals have a mean and median values of 2.54 and 2.1, respectively. The corresponding values for non-MC intervals are 1.12 and 0.7, respectively. Clearly, there is enhancement in both MCs and non-MC events, but higher O7O6 ratios are found for MCs. When we consider event-averaged O7O6 values, we see that the mean and median values are still above the threshold for MCs, but slightly below for non-MC events. This may be due to the fact that the O7O6 values have time structure within the ICME interval (see Figure 1), which might have caused smaller O7O6 when averaged over the event. Comparing the QFe and O7O6 values, we see that QFe is a better indicator of ICMEs than O7O6.

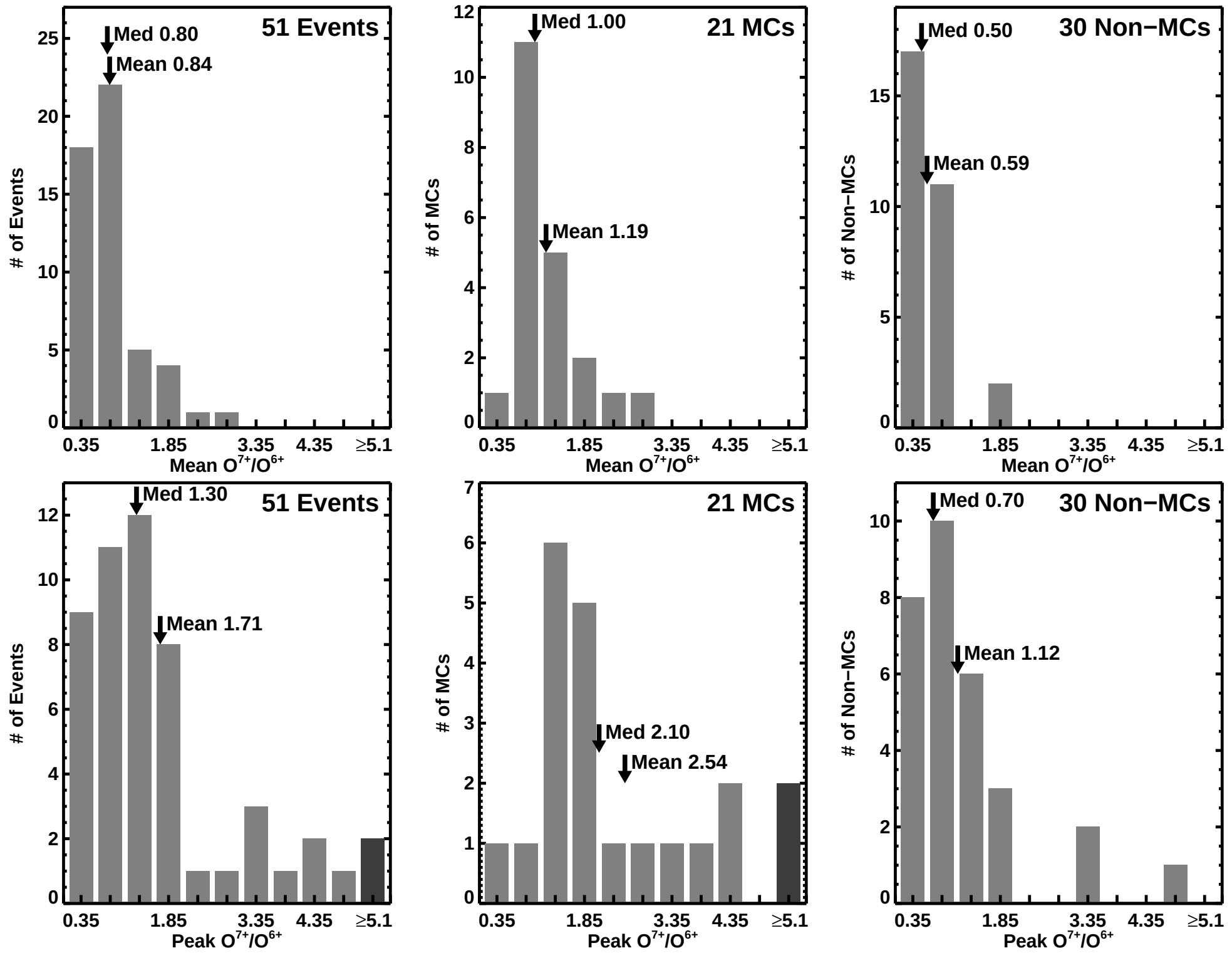

Figure 3. Distribution of O7O6 inside ICMEs using average (top) and peak (bottom) values within the ICME intervals. MCs and non-MC ICMEs distinguished. The mean and median values shown on the plots.

### 3.2 Charge State and ICME Durations

The ICME boundaries given in Table 1 were obtained from plasma and magnetic (plasmag) signatures. In order to check the durations of ICME events from the charge states alone, we measured the duration when QFe and O7O6 remained above the threshold values ignoring the ICME ending time. In other words, if the charge state remained above the threshold, we counted the duration until the values dropped to the threshold values. In some cases the value never came down, so the end time is the end time of the data set. The distributions in Figure 4 show that the mean and median plasmag durations are 16.5 and 16.9 h, respectively for all the ICMEs. The MC and non-MC durations taken separately are not substantially different from these values. However, when QFe is used (middle panel of Figure 4), the MC distribution gets much wider and the mean and median values are substantially higher (34.5 and 37.7 h, respectively). The O7O6 values also had a wider distribution (bottom panels of Figure 4), but to a less extent (mean and median O7O6 values: 23 and 27.7 h, respectively). In non-MC events, the plasmag and QFe durations were similar, whereas the O7O6 durations were slightly smaller. One problem with these durations is that we have not paid attention to the solar wind structure beyond the rear boundary of the ICMEs. The charge state enhancement may be due to poor definition of the boundaries from plasmag signatures or due to weaker ICMEs that follow the ICME in question.

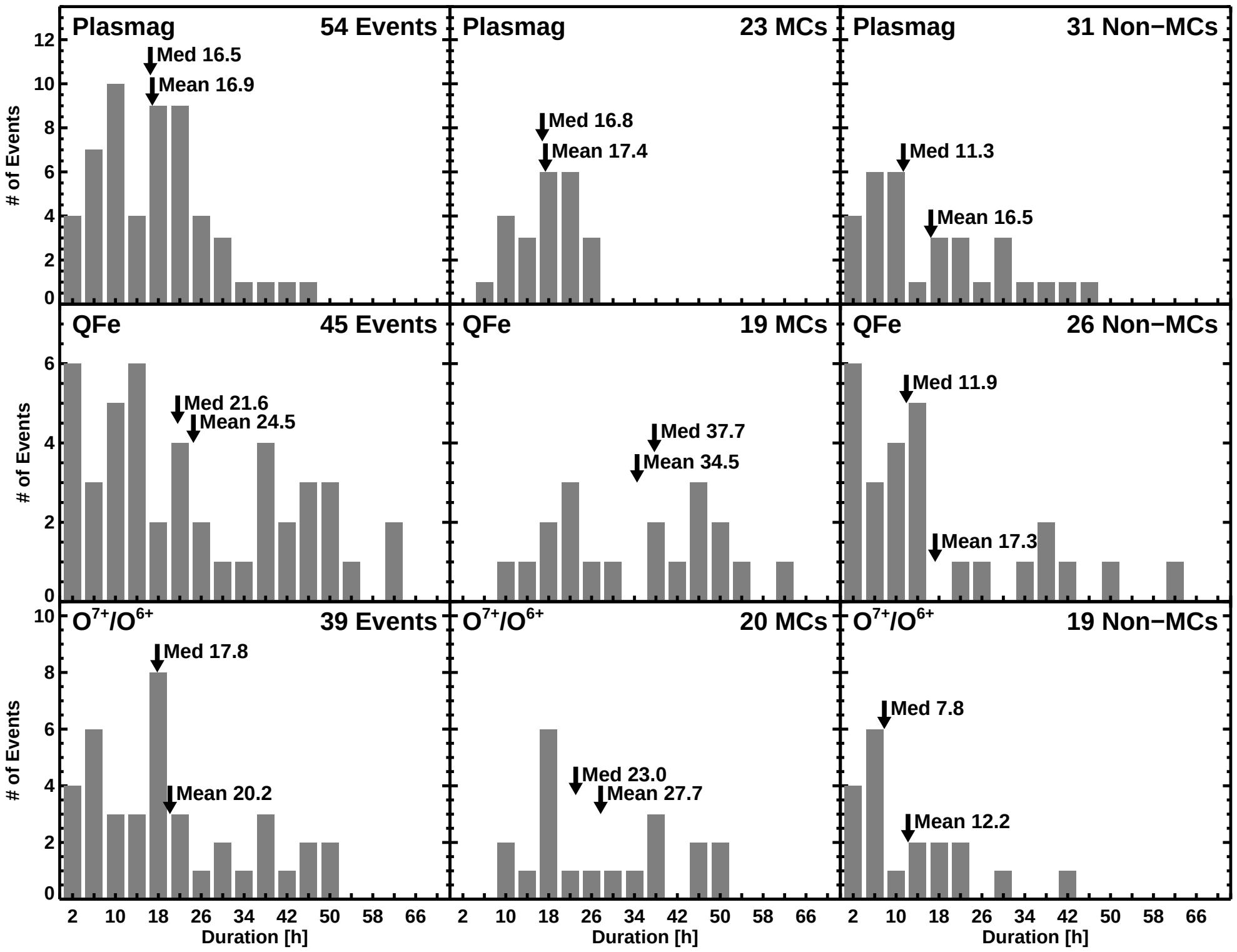

Figure 4. ICME durations based on plasma signatures (top - plasmag), QFe (middle), and O7O6 (bottom). The mean and median durations (in hours) are noted on the plots.

We have also not considered the fact that the ICME interval may contain subintervals of low charge states from prominence material (Burlaga *et al.* 1998; Gopalswamy *et al.* 1998; Lepri and Zurbuchen2010; Gilbert *et al.* 2012). In order to avoid the uncertainty on the ICME signatures outside of the plasmag boundaries, we computed the duration within the plasma ICME boundaries, by summing up only those subintervals when the charge states remained above the threshold values. As Figure 1 shows, in the 2000 July 11 event, the plasmag duration of EJ2 is~27.6 h, whereas the QFe and O7O6 values remain above the threshold only for 24 and 21 h, respectively. The reduction is essentially due to time structure in the charge state profiles (especially for O7O6). This suggests that the ICME may not be uniformly filled with hot plasma, but in patches as in Figure 1 (EJ2). Numerical simulations also suggest such spatial inhomogeneity within the CME flux rope (see *e.g.*, Lynch *et al.* 2011). Figure 5 shows the distributions of these reduced durations. Now, the QFe and O7O6 enhancements have similar durations that are substantially below the plasmag durations given in Figure 4. Just by comparing the mean values, we see that the charge state durations constitute a fraction of the plasmag duration in the range 0.56 to 0.74. Taking the average durations in columns 23 and 28 in Table 1, we see that the ICMEs are filled with 67% enhanced QFe and 63% enhanced O7O6. This suggests that the hot plasma is filling only part of the CMEs when they are released near the Sun. Furthermore, both the QFe and O7O6 durations in MCs are generally longer than those in non-MC events. This is significant because this may be related to the fact that the observing spacecraft may not be passing through the nose of the ICME in the case of non-MC events thereby intercepting less number of patches of high charge state. Such an interpretation would be consistent with the non-radial motion of the CMEs that result in non-MC ICMEs.

### 3.3 Flare comparisons

Since the flare heating is ultimately responsible for the injection of hot plasma into the CMEs, it is imperative that we compare the flare properties of the MC and non-MC events. Figure 6 shows the flare size distributions for MC, non-MC, and the combined set. The mean and median flare size of flares associated with the ICMEs in general fall in the M class suggesting that most of the flares are major ones. When MC and non-MC events are considered separately, we see that the flares of the non-MC events are slightly smaller in size. For MCs, the median size remains in M class whereas it is in C class for the non-MC events. The mean sizes are higher than the median sizes because of the asymmetry, but even there the MC-associated flares are one class higher. Thus there is some indication that we are dealing with slightly weaker flares in the case of non-MC events, although there is a heavy overlap in flare sizes between the two populations. What is really needed in the flare is that the plasma temperature should reach sufficient level to ionize enough number of ions to be detected as a charge state enhancement at 1 AU. To see this, we used the soft X-ray intensities in the two GOES energy channels to obtain the flare temperature. We were able to determine the flare temperature for 22 MC events. There were several weak events identified as eruptive prominence (EP) event or an event with weak post-eruption arcade (PEA). The solar source of one of the CMEs is an eruptive prominence (EP) event (2000 August 11). The others were non-MC events with low soft X-ray flux that we were not able to determine the flare temperature. We discuss these weak events separately in a later subsection.

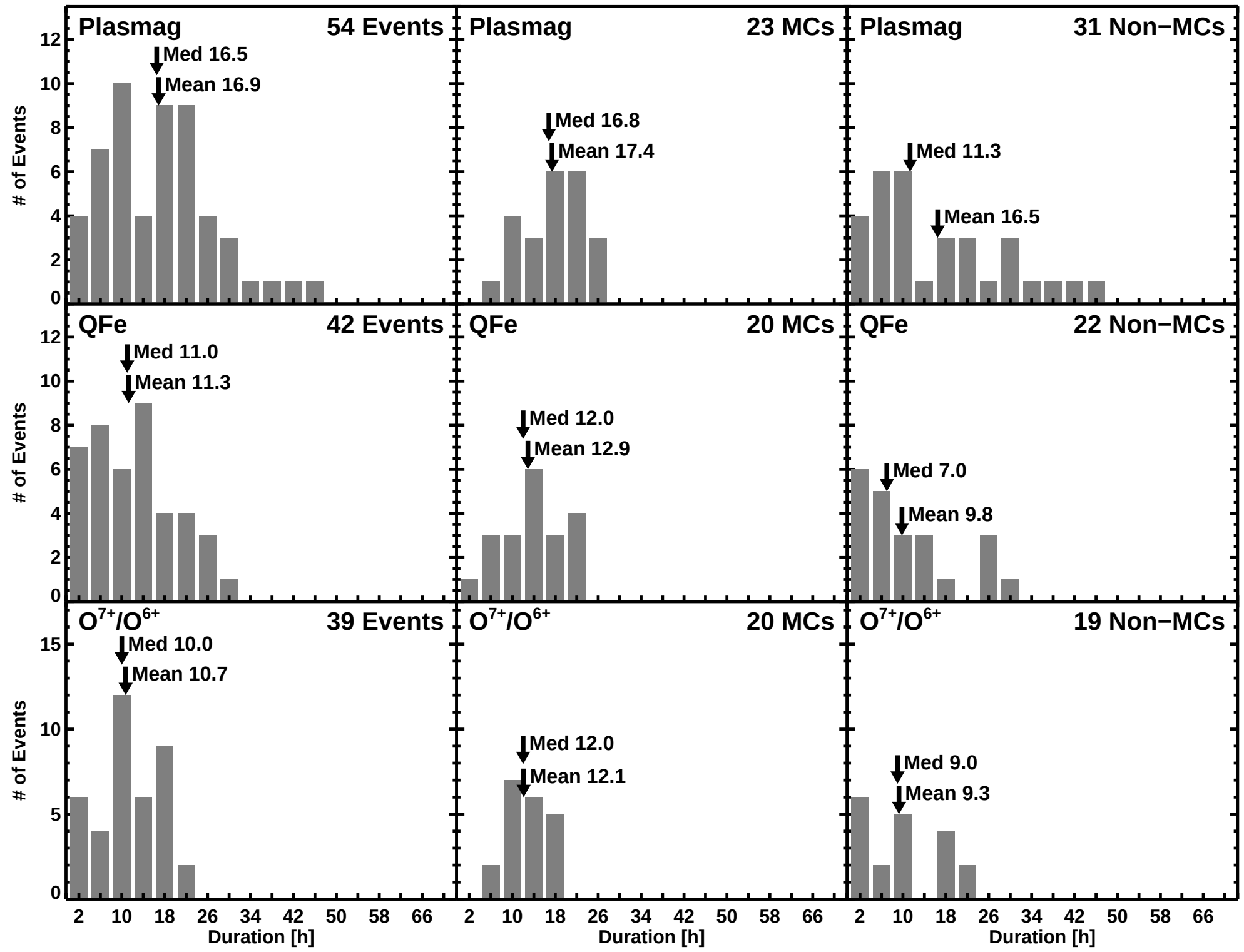

Figure 5. ICME durations based on plasmag signatures (top) compared with reduced durations obtained from QFe (middle), and O7O6 (bottom) signatures. The mean and median durations (in hours) are noted on the plots.

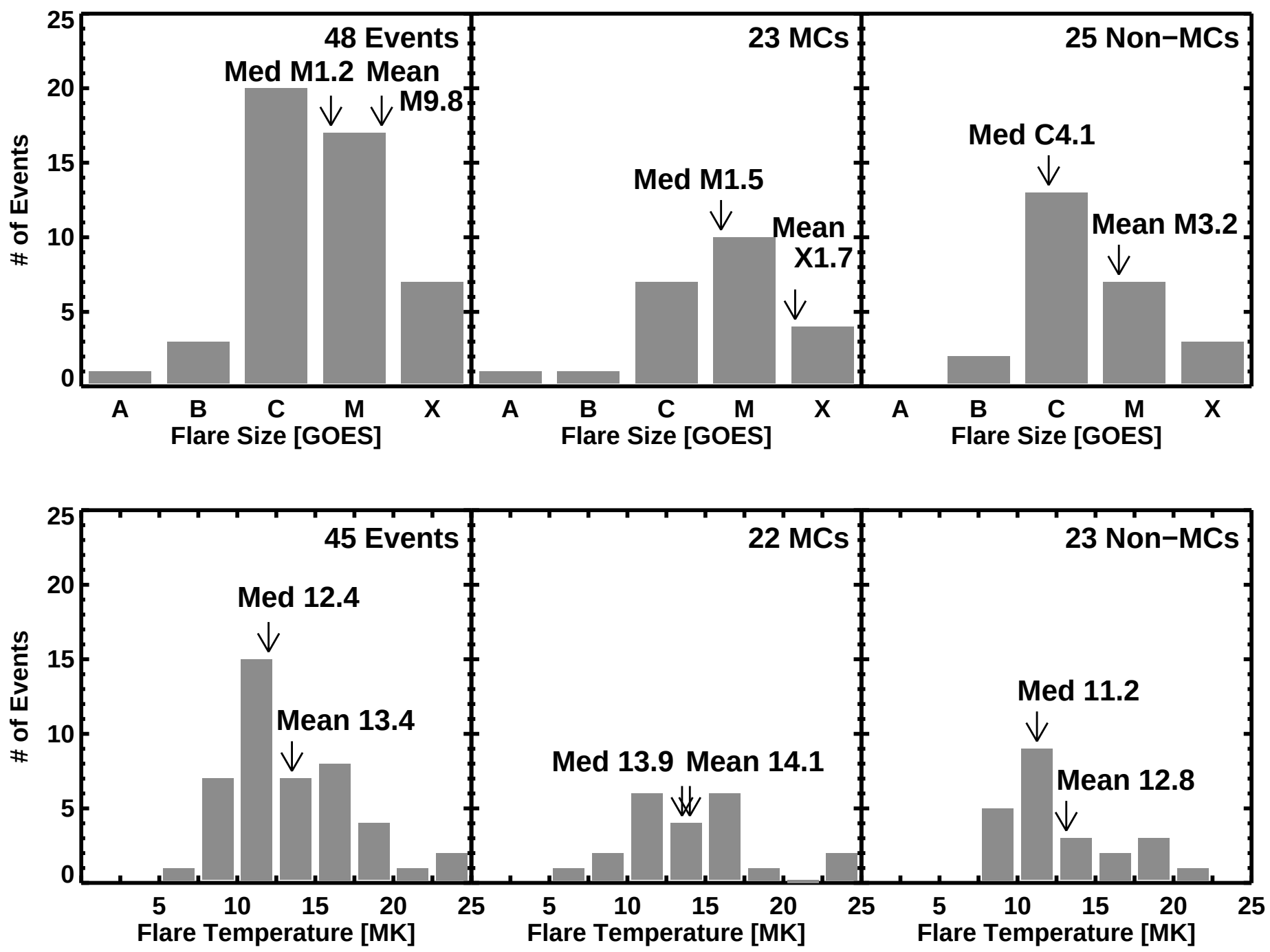

Figure 6. Size and temperature distributions of GOES soft X-ray flares for the selected ICME events with the flares associated with MCs and non-MC MCs distinguished. The mean and medium values of the distributions are marked on the plots. For nine events, the soft X-ray intensity was too low to calculate the temperature.

Figure 6 also shows the flare temperature distributions for 22 MCs and 23 non-MC events. The flare temperatures range from 5 MK to 25 MK. The mean and median flare temperatures are nearly the same for both MC and EJ events. The range of temperatures is more than adequate in producing the observed QFe and O7O6 enhancements (Bame *et al.* 1979; Lepri *et al.* 2001). Thus we conclude that the flares involved in both MC and non-MC events have similar flare sizes and temperatures, suggesting that the availability of hot plasmas is about the same for the two populations.

### 3.3.1 Correlation between flare size, flare temperature and charge states

Reinard (2005, 2008) reported a general increase in charge state ratios as a function of the flare size. She grouped the flares into C, M, and X classes and found that both O7O6 and QFe values were enhanced the greatest in the case of X flares and the least in the case of C-class flares. In our sample, we have even A and B class flares, so we use scatter plots between the flare size and temperatures on the one hand and the charge states on the other. For the eight EP events, there is no flare information available, so we have not used them. Excluding events with data gaps, we have 20 MC and 23 non-MC events for which we show the scatter plots in Figure 7 between the flare intensity and the peak and event-averaged QFe values. The high degree of overlap between the MC and non-MC data points is quite obvious. There is definitely a positive correlation between QFe and flare size for all the three cases shown: MC events, non-MC events and for the combined set. The correlation coefficient (r) is 0.5 for the peak QFe within the ICME interval. The probability (p) that the observed correlation is by chance is very low: $4.9 \times 10^{-4}$. When the event-averaged QFe is used, the correlation is even better (r = 0.59) with p = $1.9 \times 10^{-5}$. The correlation coefficient is reasonably high for MC events (r = 0.56 for peak QFe and 0.61 for the averaged QFe with p values of $9.0 \times 10^{-3}$ and $3.2 \times 10^{-3}$, respectively). For the non-MC events, the correlation is somewhat weaker (r =0.31 with p = 0.16 for peak QFe and r = 0.46 with p = 0.027 for event-averaged QFe). In Figure 7 we see some outliers at low values of QFe. These outliers could be due to incomplete heating of prominence material or merely because the spacecraft observations did not sample the portion of the ICME that contained enhancements (due to geometrical constraints, in-situ observations of charge state enhancements provide only a lower limit on the initial heating). When the outliers at the bottom of the plot are excluded, the correlation improves significantly: For peak QFe, the correlation coefficients are 0.60 (p = $2.6 \times 10^{-5}$, combined set), 0.58 ($7.4 \times 10^{-3}$, MC events), and 0.49 (p = 0.023, non-MC events). For event-averaged QFe, the correlation is even better: 0.68 (p = $6.8 \times 10^{-7}$, combined set), 0.63 ($3.2 \times 10^{-3}$, MC events), and 0.63 (p = $1.7 \times 10^{-3}$, non-MC events). The correlation analysis confirms the flare-size dependence of QFe. Furthermore, the high overlap between the data points from MC and non-MC events suggests that they should be similar objects.

The correlation analysis done for O7O6 values against flare size are shown in Figure 8. One can see significant overlap between MC and non-MC events, but the non-MC events are generally concentrated toward the lower charge state values as we also showed using the distributions in Figure 3. This is particularly clear in the event-averaged O7O6 values shown in the right side panel of Figure 3.

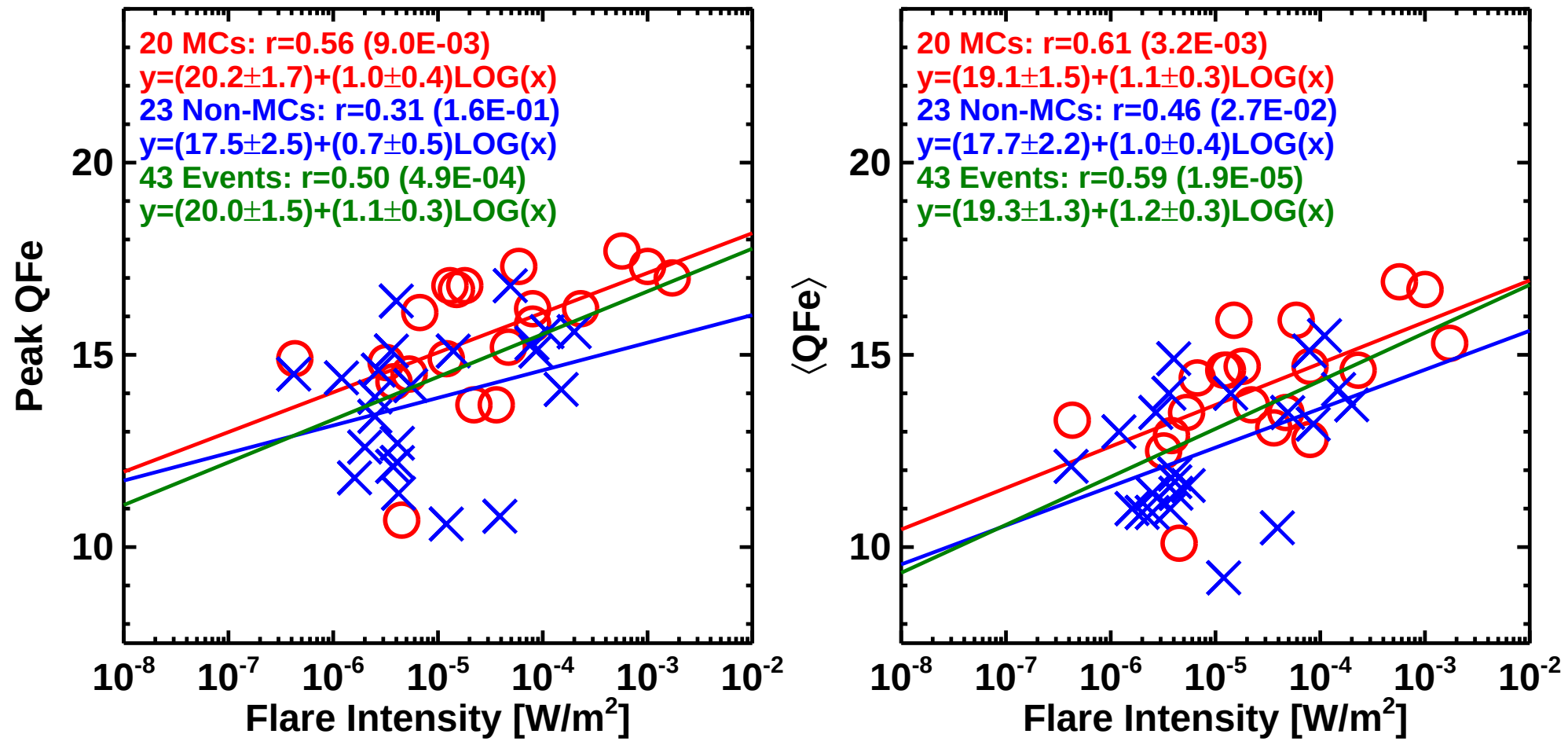

Figure 7. Scatter plots between the soft X-ray flare size and the peak (left) and average (right) QFe in ICMEs. MCs and non-MCs are denoted by circles and crosses, respectively. The correlation coefficients (r) and the regression lines for the MC and non-MC events as well as the combined set (43 events) are shown on the plots. The probability of obtaining the correlation by chance is indicated in parentheses.

Even though the correlation is positive, it is much weaker compared to the QFe – flare size correlation. For the combined set, the correlation coefficients are similar for peak (r =0.42 with p = 7. $0\times10^{-3}$) and event-averaged (0.4 with p = 6. $5\times10^{-3}$) O7O6. The correlation is still reasonable for MC events: r = 0.4 (p = 0.08) and 0.32 (p = 0.15) for peak and event-averaged O7O6, respectively. The lowest correlation is for the non-MC events: 0.24 (p = 0.25, peak O7O6), 0.25 (p = 0.23, event-averaged O7O6). Note that the p values are high indicating low confidence levels (75% and 77%) for the peak and event-averaged O7O6 vales.

The correlations of charge states measures with flare temperature are similar to those with the peak soft X-ray flux. The correlation coefficients and the p-values shown in Figures 9 and 10 indicate that all the correlations are highly significant, confirming the importance of flares in creating the high charge states observed inside ICMEs of both types. The lowest correlation obtained is for peak O7O6 in non-MC events: r = 0.33 with p=0.13. The confidence level of this correlation is only 87%.

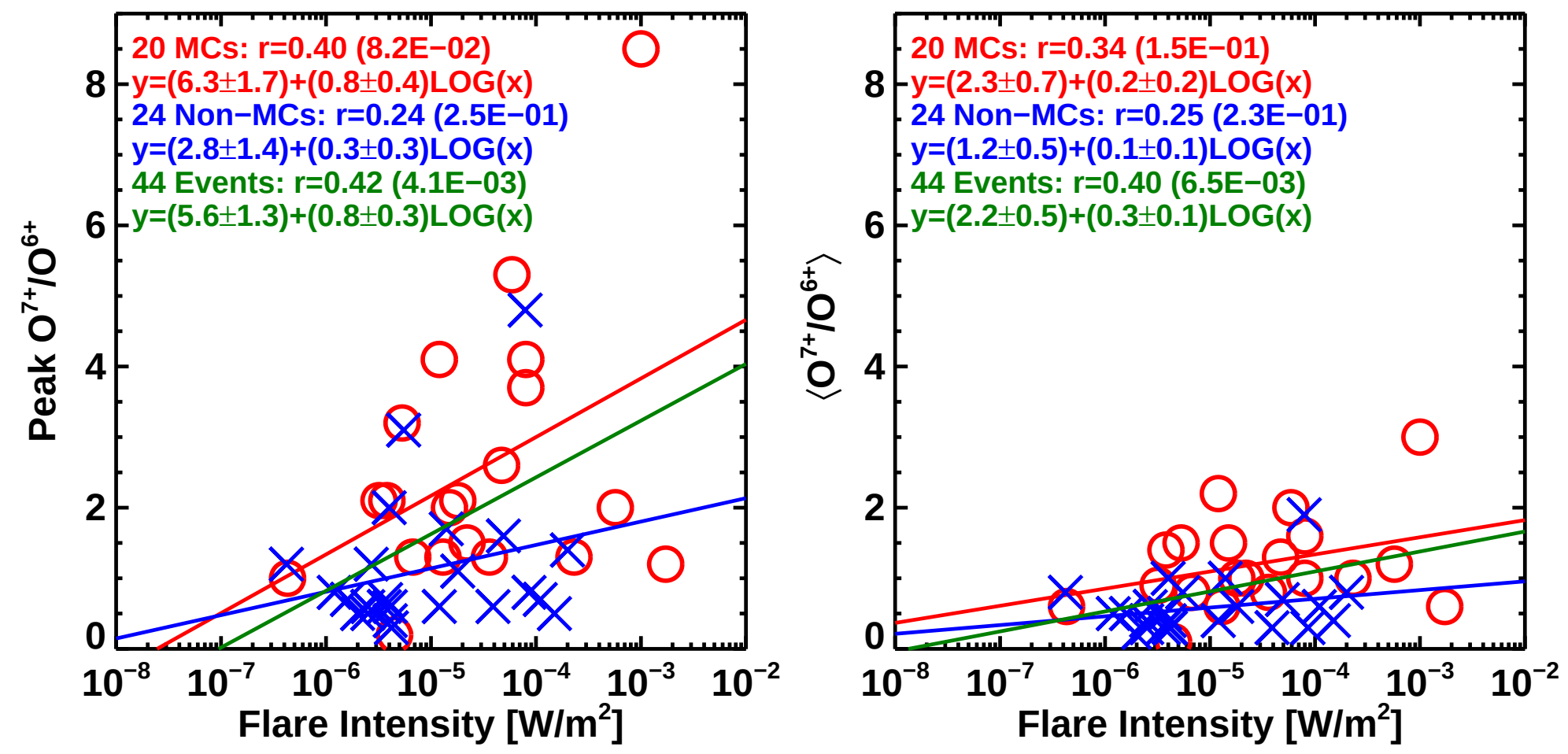

Figure 8. Scatter plots between the soft X-ray flare size and the peak (left) and average (right) O7/O6 ratios in ICMEs. MCs and non-MC are denoted by circles and crosses, respectively. The correlation coefficients (r) and the regression lines for the MC and non-MC events as well as the combined set (44 events) are shown on the plots. The probability of obtaining the correlation by chance is indicated in parentheses.

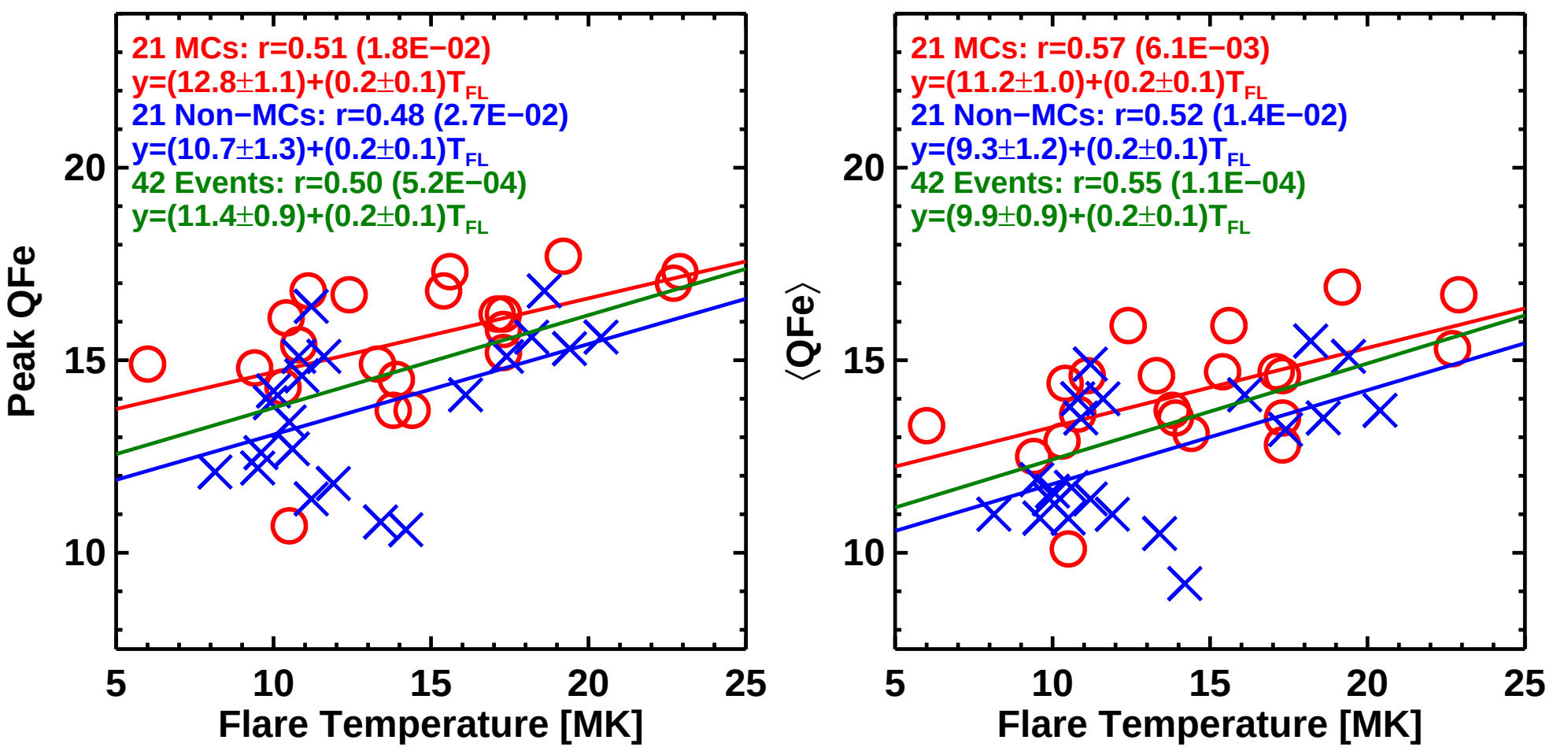

Figure 9. Scatter plots between flare temperature and the peak (left) and average (right) QFe in ICMEs. MCs and non-MCs are denoted by circles and crosses, respectively. The correlation coefficients (r) and the regression lines for the MC and non-MC events as well as the combined set (42 events) are shown on the plots. The probability of obtaining the correlation by chance is indicated in parentheses.

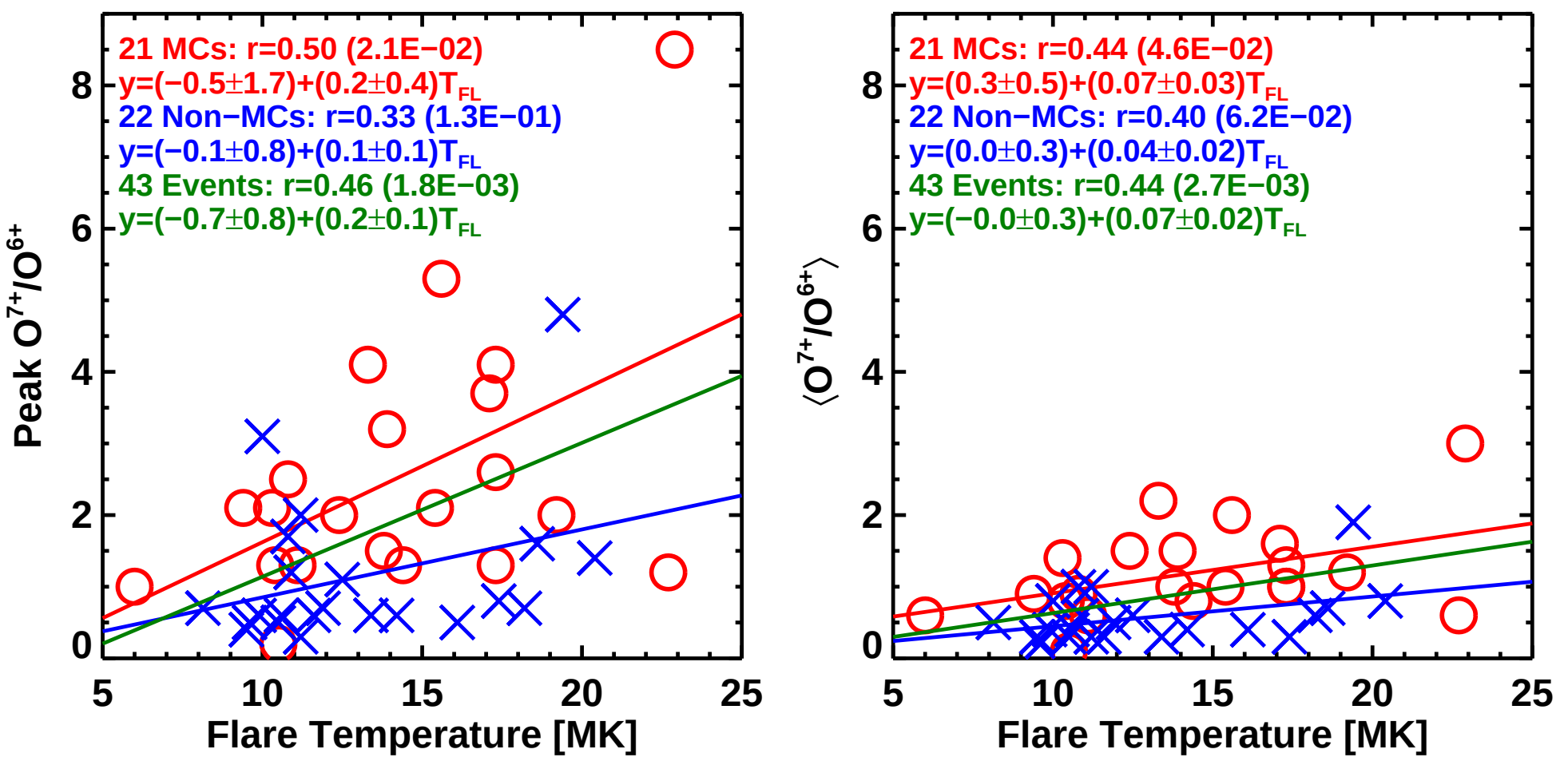

Figure 10. Scatter plots between flare temperature and the peak (left) and average (right) O7/O6 ratio in ICMEs. MCs and non-MCs are denoted by circles and crosses, respectively. The correlation coefficients (r) and the regression lines for the MC and non-MC events as well as the combined set (43 events) are shown on the plots. The probability of obtaining the correlation by chance is indicated in parentheses.

### 3.4 CME comparisons

We have seen in the previous sections that there is no significant difference between flares associated with the MC and non-MC events. The flare signatures are contained within the CME as the charge state enhancements. Is there any characteristic difference between the CMEs associated with the two types of ICMEs? In order to check this we have plotted the speed, width, and acceleration distributions of the MC and non-MC events in Figure 11. The speeds of white-light CMEs near the Sun are about two times larger than the average speed of the general population of CMEs. The speeds of MC associated CMEs (mean 934 km $s^{-1}$) are similar to speeds (mean 782 km $s^{-1}$) reported before without the longitude restriction (Gopalswamy, Yashiro, and Akiyama, 2007; Gopalswamy *et al.* 2010b). This is because the solar sources of MC-associated CMEs tend to be closer to the disc center. On the other hand, the solar sources of non-MC ICMEs are generally at larger distances from the central meridian, so their speed measurement is subject to less projection effects. Accordingly, the average speed of CMEs associated with non-MC events is somewhat higher (955 km $s^{-1}$ *vs*. 772 km $s^{-1}$). The events in Figure 11 are both from disk center, and hence subject to similar projection effects resulting in similar speeds.

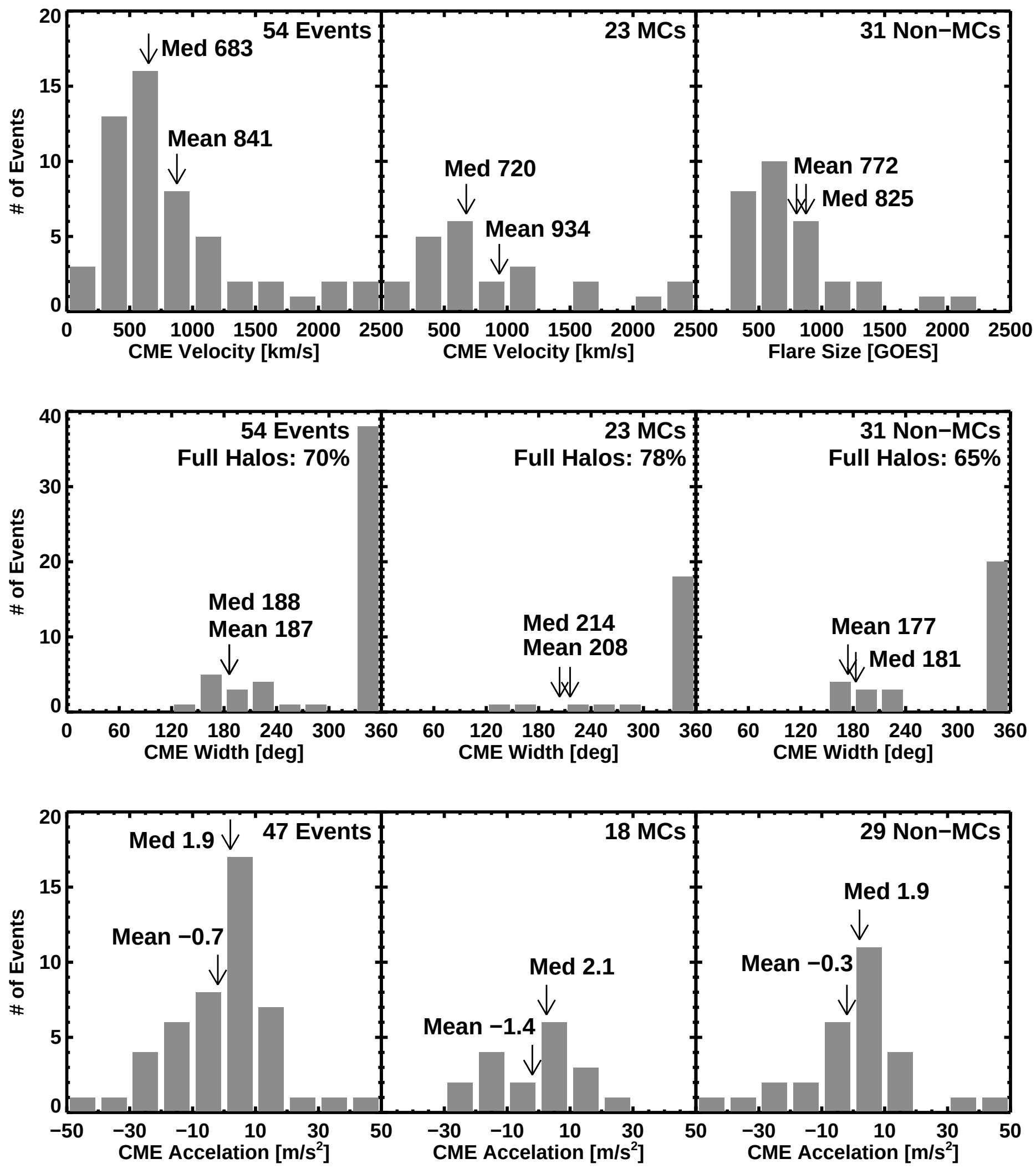

Figure 11. Speed, width and acceleration of CMEs associated with the ICMEs in question. MC and non-MC values are compared with each other and with the combined set. In the width distributions, the fraction of halo CMEs is indicated.

Such high speed CMEs from the disk center are expected to appear as halo CMEs in the coronagraphic field of view. The width of the halo CMEs is not known, but measurements of limb CMEs reveal that faster CMEs are generally wider (Gopalswamy *et al.* 2009d). Again wider CMEs are more massive (Gopalswamy *et al.* 2005), indicating that faster CMEs are generally more energetic. In other words, halo CMEs are expected to be generally more energetic. In fact, the fraction of halo CMEs in a population is an indicator of the average energy of the population: higher the halo fraction, larger is the kinetic energy. Gopalswamy *et al.* (2010b) found that the majority of CMEs arriving at Earth are halos: 59% of CMEs associated with MCs and 60% associated with non-MCs. Figure 11 shows even a larger fraction of halo CMEs (70% for MC+non-MC events) in the present study because they originate closer to the disc center compared to all halos. The halo fraction is the highest with 76% for MC-associated CMEs, while somewhat smaller (65%) for the non-MC CMEs. A CME needs to be relatively fast to become a halo CME when it originates farther from the disk center (Gopalswamy *et al.* 2010c).

The acceleration measurement is generally difficult and is accurate only for slow CMEs from the limb: because there are no projection effects for limb CMEs and many data points can be obtained for slow CMEs. The CMEs in question are subject to projection effects because they all come from close to the disk center. Fortunately, comparing the acceleration of MC and non-MC CMEs is possible because both sets are subject to similar projection effects. We see from Figure 11 that the accelerations are similar for MC, non-MC, and the combined set. Gopalswamy (2010) showed that for a large number of limb CMEs, the mean acceleration was -3.1 m $s^{-2}$, which is only slightly larger than the mean values in Figure 11. One small difference is that the distribution peaks in the 0-10 m $s^{-2}$ bin. A closer examination of these events in this bin reveals that most of these CMEs are radio quiet. i.e., they did not produce a type II radio burst anywhere between the Sun and Earth, even though they were associated with IP shocks at 1 AU. Accelerating CMEs become fast enough to drive shocks generally far away from the Sun (beyond 10 Rs), so they either produce type II bursts at kilometric wavelengths (Gopalswamy, 2006b) or none at all (Gopalswamy *et al.* 2010a). When we examined the type II burst association of the 54 events, we found that 17 were radio quiet (no type II burst association). The vast majority of the radio-quiet CMEs are non-MC events (14 *vs*. 3 MCs), consistent with the positive acceleration bias seen in Figure 11. Only 4 of the 14 RQ CMEs associated with non-MC events were decelerating.

In summary, we see that the basic properties of CMEs (speed, width, and acceleration) in the MC and non-MC events are very similar. The only exception we find is a slightly larger number of radio-quiet CMEs among the non-MC events (14 out of 31 non-MC events or 45% are radio quiet, while 3 out of 23 MC events or 13% are radio quiet). Now let us look at the correlation between CME speed and charge state measures.

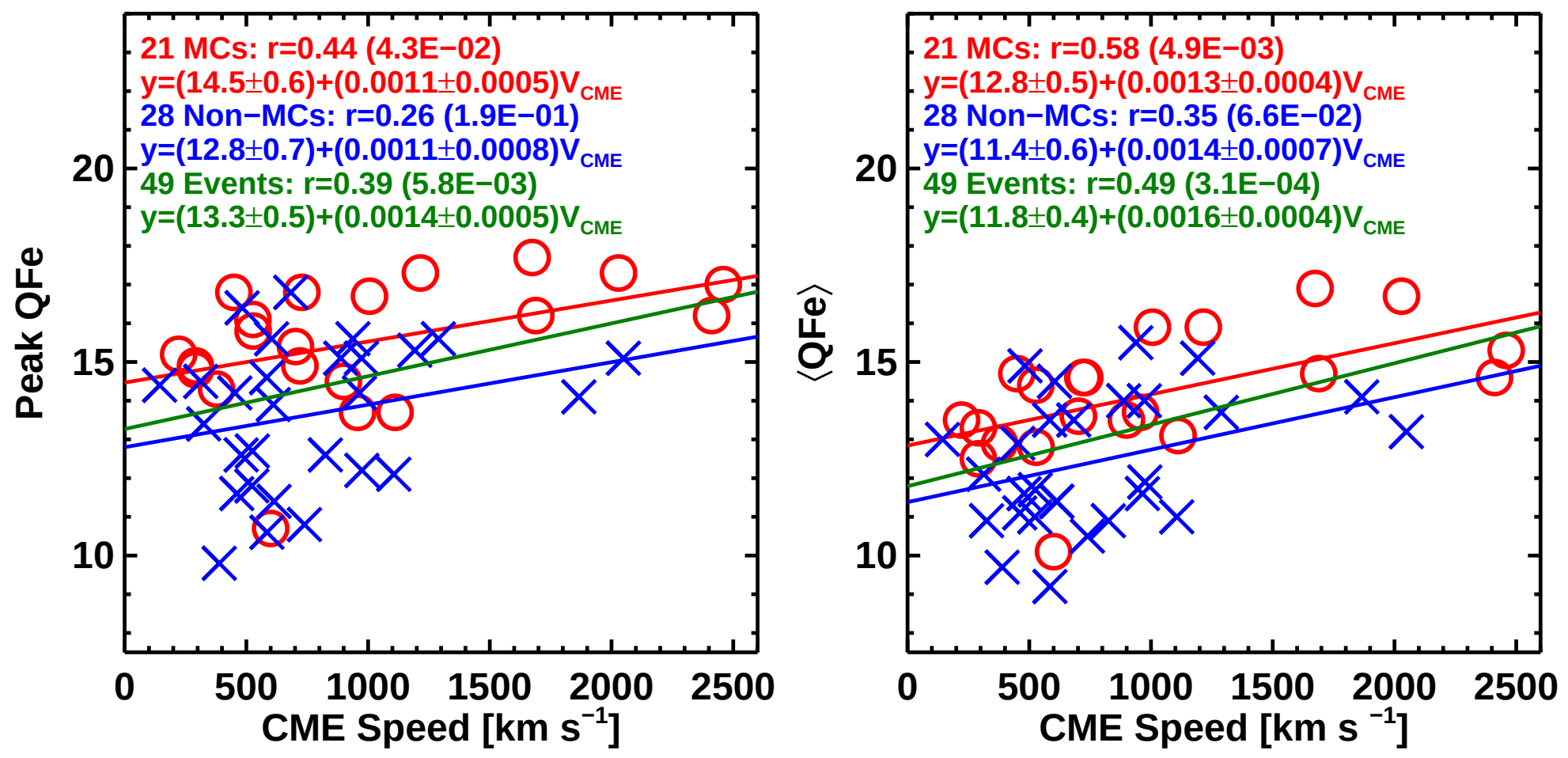

Figure 12. Scatter plots between CME speed and the peak (left) and average (right) QFe in ICMEs. MCs and non-MCs are denoted by circles and crosses, respectively. The correlation coefficients (r) and the regression lines for the MC and non-MC events as well as the combined set (49 events) are shown on the plots. The probability of obtaining the correlation by chance is indicated in parentheses.

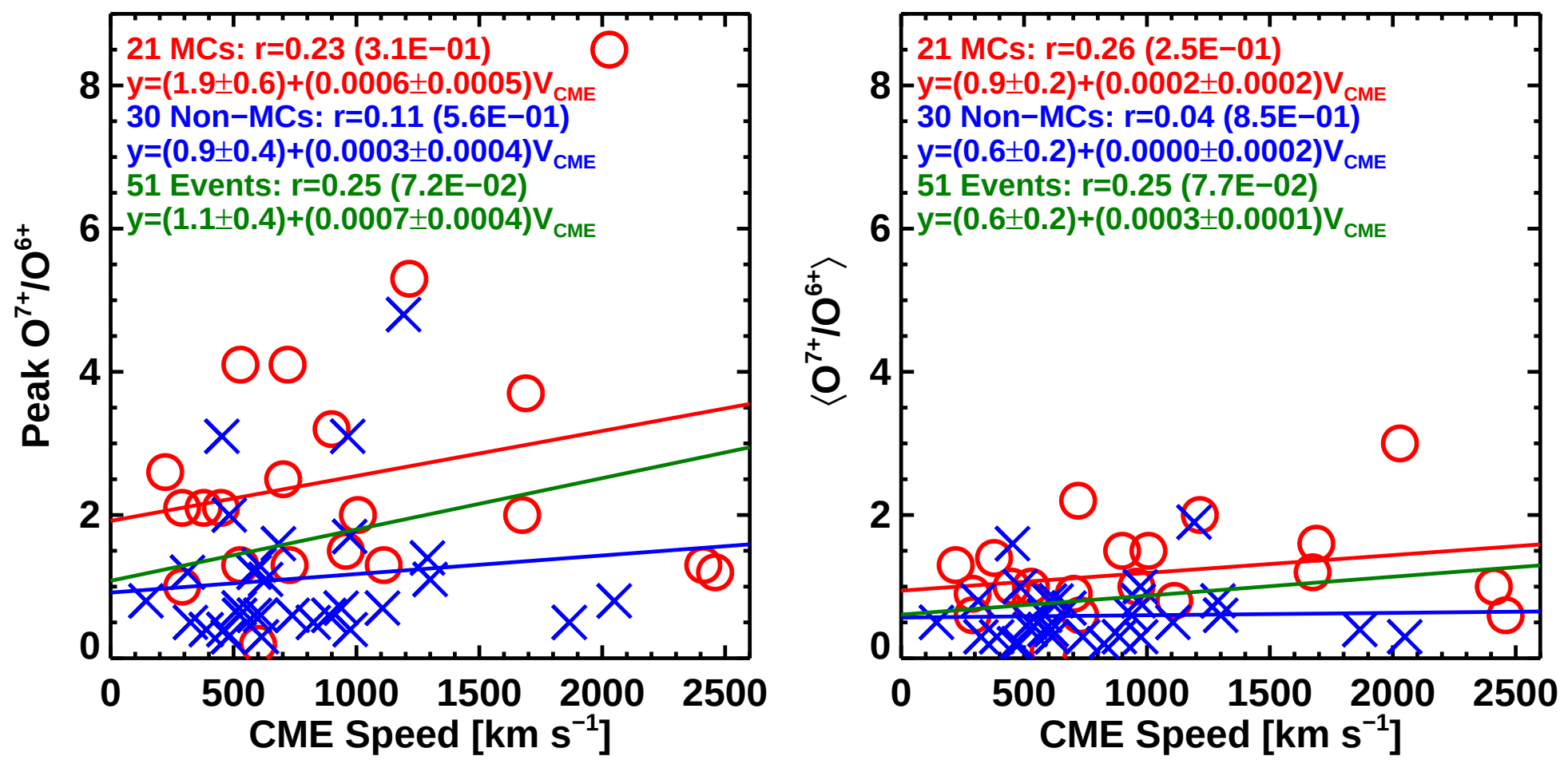

Figure 13. Scatter plots between CME speed and the peak (left) and average (right) QFe in ICMEs. MCs and non-MC are denoted by circles and crosses, respectively. The correlation coefficients (r) and the regression lines for the MC and non-MC events as well as the combined set (49 events) are shown on the plots. The probability of obtaining the correlation by chance is indicated in parentheses.

There is generally a positive correlation between the CME speed and QFe. Figure 12 shows that the correlation coefficients range from 0.26 to 0.58. The weakest correlation ($r = 0.26$) is for peak QFe with $p = 0.19$ indication that the confidence level is only 81%. On the other hand the CME speed is poorly correlated with O7O6 values as can be seen in Figure 13 We think the CME speed – charge state correlation essentially reflects the correlation between CME speed and flare size (see *e.g.*, Gopalswamy, 2010) because CMEs do not play any role in the creation of high charge states.

**Table 2** Correlation coefficients for flare/CME properties and QFe and O7O6

| | Correlation coefficient for QFe[d] | | | Correlation coefficient for O7/O6[d] | | |
|---|---|---|---|---|---|---|
| | MC | Non-MC | MC+non-MC | MC | non-MC | MC+non-MC |
| Flare Size | 0.56 (0.9%)<br>0.61 (0.3%) | 0.31 (16%)<br>0.46 (2.7%) | 0.50 (0.05%)<br>0.59 (0.002%) | 0.40 (8.2%)<br>0.34 (15%) | 0.24 (25%)<br>0.25 (23%) | 0.42 (0.4%)<br>0.40 (0.7%) |
| Flare Size XO[a] | 0.58 (0.7%)<br>0.63 (0.3%) | 0.49 (2.3%)<br>0.63 (0.2%) | 0.60 (0.003%)<br>0.68 (0.00006%) | 0.16 (51%)<br>0.11 (66%) | 0.24 (25%)<br>0.07 (75%) | 0.29 (6.1%)<br>0.25 (12%) |
| Flare T[b] | 0.51 (1.8%)<br>0.57(0.6%) | 0.48 (2.7%)<br>0.52 (1.4%) | 0.50 (0.05%)<br>0.55 (0.01%) | 0.50 (2.1%)<br>0.44 (4.6%) | 0.33 (13%)<br>0.40 (6.2%) | 0.46 (0.2%)<br>0.44 (0.4%) |
| Flare T XO[a] | 0.51 (1.9%)<br>0.57 (0.8%) | 0.67 (0.2%)<br>0.66 (0.2%) | 0.59 (0.008%)<br>0.62 (0.002%) | 0.27 (26%)<br>0.44 (4.6%) | 0.26 (27%)<br>0.17 (47%) | 0.28 (7.5%)<br>0.41 (0.7%) |
| CME V[c] | 0.44 (4.3%)<br>0.58 (0.5%) | 0.26 (19%)<br>0.35 (6.6%) | 0.39 (0.6%)<br>0.49 (0.03%) | 0.23 (31%)<br>0.26 (25%) | 0.11 (56%)<br>0.04 (84%) | 0.25 (7.2%)<br>0.25 (7.7%) |
| CME V XO[a] | 0.49 (2.8%)<br>0.63 (0.2%) | 0.26 (19%)<br>0.35 (6.6%) | 0.38 (0.7%)<br>0.49 (0.04%) | -0.04 (87%)<br>0.26 (25%) | 0.03 (90%)<br>-0.01 (96%) | 0.08 (60%)<br>0.27 (6.7%) |

[a]XO indicates that a few outliers were excluded; [b]Flare temperature derived from GOES soft X-ray intensities; [c]Speed of white-light CMEs from LASCO; The upper (lower) entries are peak (average) charge state values within the ICME interval. The percentage values in parentheses denote the probability that the observed correlation is due to chance. Smaller is this probability, the higher is the confidence level in the reality of the correlation.

Table 2 summarizes various correlation coefficients discussed above for QFe and O7O6. The probability that a correlation is by chance is given by the number in parentheses. Any p value more a few percent is an indication that we have low confidence in the correlation. The confidence level is roughly 1-p. We have listed the correlation of QFe and O7O6 with flare intensity, flare temperature, and CME speed. We have also listed the correlation coefficients obtained by eliminating a few outliers. These cases are denoted by the “XO” (for excluding outliers). Barring one or two cases, the charge states have generally a high correlation for QFe. On the other hand, O7O6 correlations are generally weaker, especially with CME speed. The poorest correlations are between O7O6 and CME speed for non-MC events. The lower correlations with CME properties are understandable because CME properties do not decide the creation of charge states.

### 3.5 Weak events

We saw that there were eight weak events in terms of flare size. These were eruptive prominence events with clear post-eruption arcades. Even though the flare signature in these events is extremely weak, the post eruption arcades (in soft X-rays, EUV, or microwaves) are very prominent. The soft X-ray flux derived from imaging observations (Yohkoh/SXT) is well below the GOES soft X-ray background level, so these events do not have flares listed in the SGD. All but three of these EP events had a charge state enhancement. The exceptions are the 2001 March 22, 2001 August 12, and 2002 May 20 events. Figure 14 shows the solar source of the 2001 March 22 non-MC event as an SXT arcade on 2001 March 19 from Yohkoh/SXT. The weak east-west arcade overlying the neutral line (see Figure 14a,b). The EIT images had a clear filament channel with only a tiny filament visible in H-alpha (not shown). Figure 14c shows that the duration of the ejecta was very small suggesting the possibility that the spacecraft passed through only the northern flank of the ICME. The presence of a coronal hole to the northeast of the eruption region (see Figure 14a) might have also deflected the CME to the south. Note that our selection criterion restricts source longitudes to $\pm 15^{o}$, but not in latitudes. Therefore, CMEs could still go north or south of Earth (especially when deflected by coronal holes) and that might be why we do not always see flux ropes. The lack of charge state enhancement in this event (see Figure 14c) is likely due to the fact that the observing spacecraft is passing through the edge of the ICME and hence might have missed the charge state enhancement. The 2001 August 12 event also did not have charge state enhancement and has a similar solar source environment. The event had a clear north-south arcade in Yohkoh/SXT and SOHO/EIT images at the western edge of a north south coronal hole. Clearly the CME was deflected to the west, away from the Sun-Earth line, consistent with a very short duration ejecta (~3 h). Therefore, it is not surprising that we do not see charge state enhancement in this event. This event was already reported as a coronal-hole deflection event (Gopalswamy *et al.*, 2004, their Figure 3). Finally, the 2002 May 20 event is also associated with the eruption of a long north-south filament. The associated CME was relatively narrow ($45^o$) in the sky plane. The solar source of this event has some ambiguity because there are other CME candidates (see Cho *et al.*, 2012).

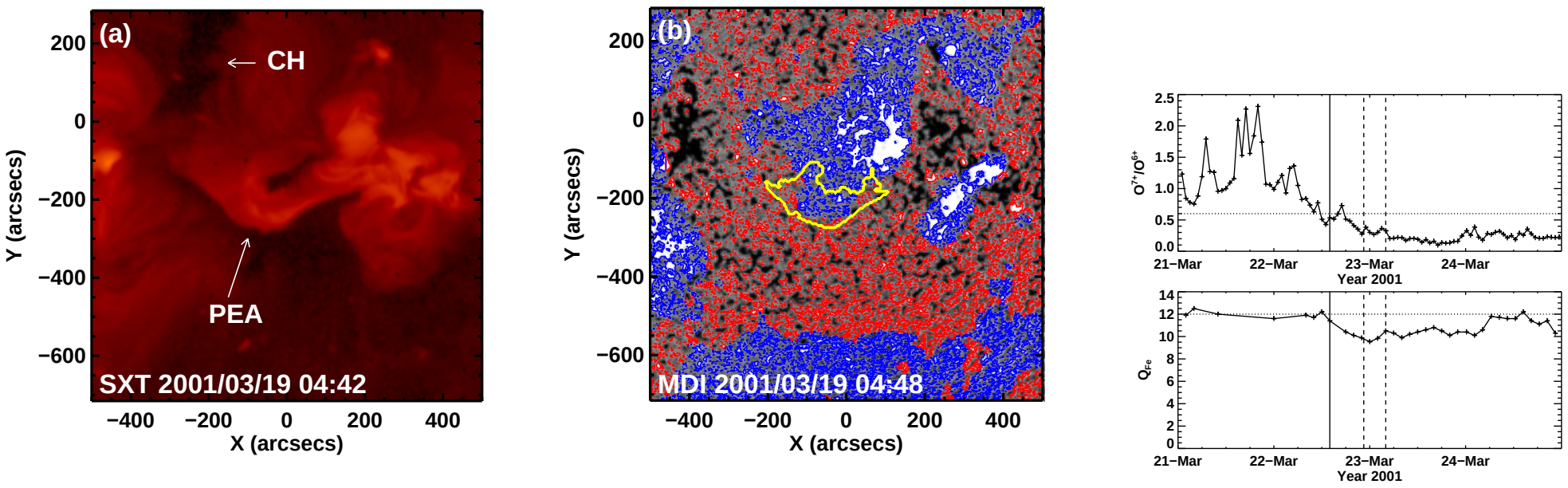

Figure 14. (a) The post-eruption arcade (PEA) as observed by Yohkoh/SXT. (b) The PEA superposed on SOHO/MDI magnetogram showing that the arcade straddles the polarity inversion line like in any eruptive event. (c) the QFe and O7O6 plots showing no charge state enhancement after the shock (vertical solid line) or during the ICME interval (marked by the vertical dashed lines).

The only EP event among the MCs is the 2000 August 10 event associated with a complex filament eruption on 2000 August 9 accompanied by a halo CME at 16:30 UT. The O7O6 ratio was ~2.5 and QFe~ 15. The arcade was observed in Yohkoh/SXT and SOHO/EIT images, but was very weak, so the event was not seen in the GOES light curve.

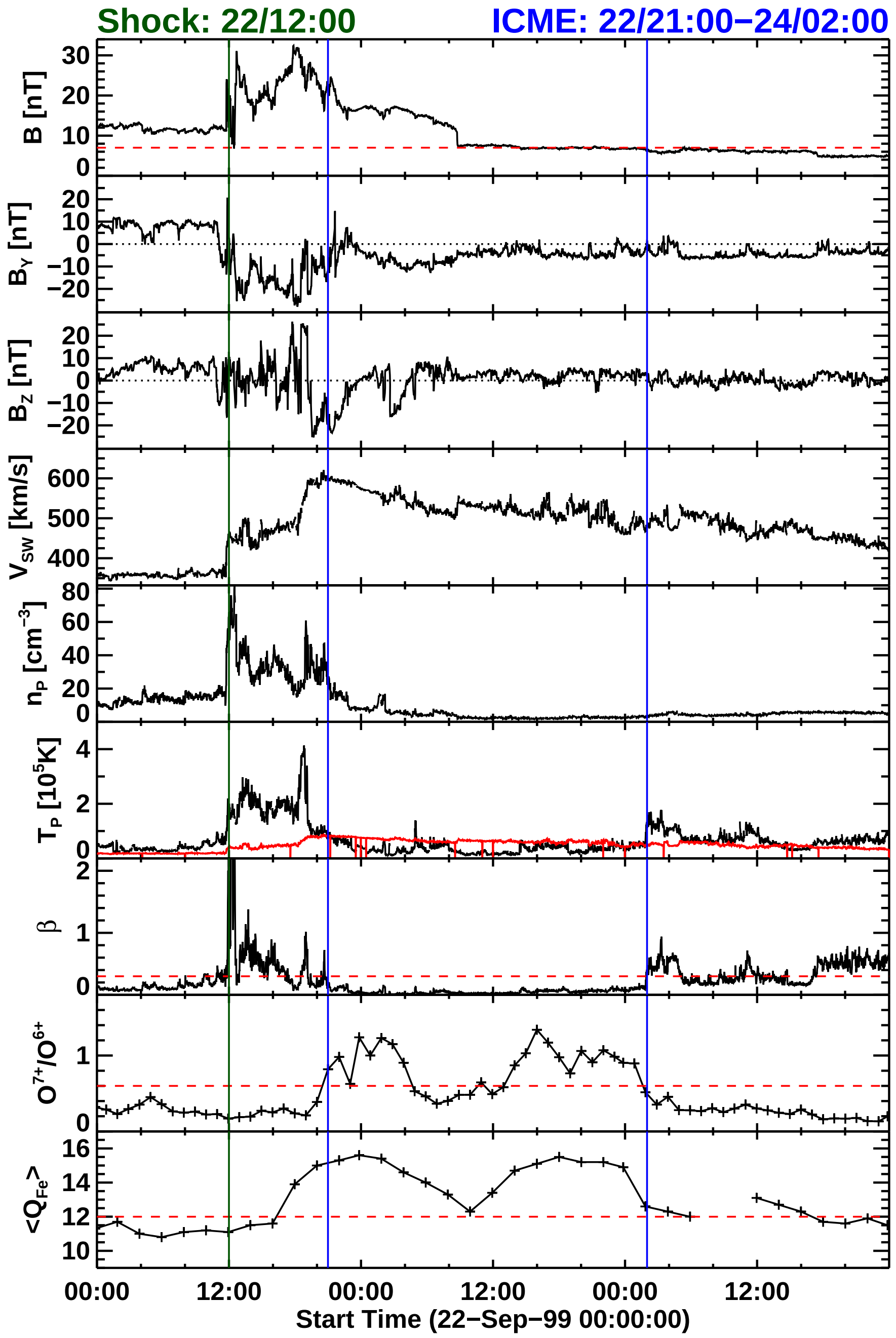

Figure 15. The O7O6 (top) and QFe (bottom) plots of the 1999 September 22 non-MC event. The solar source is identified with a filament eruption on 1999 September 19. The ejecta boundaries from the plasmag signatures are shown by the vertical dashed lines. The shock is denoted by the vertical solid line.

One of the characteristics of the EP events is that the flare structure is extremely weak, so the question arises whether high charge states can be produced in such weak flares. We already saw that 4 of the 7 events did have enhanced charge states, and in two other events, the spacecraft might have missed the flux rope. How do we reconcile these observations? In order to do this we perform a case study of the 1999 September 22 non-MC event (see Figure 15). Both QFe and O7O6 plots show a double structure, similar to many of the EP events. The O7O6 boundaries above the threshold value of 0.6 coincide well with the boundaries derived from plasmag signatures. However, the QFe signature starts 2-3 hours earlier. The peak (event-averaged) QFe and O7O6 values are 15.6 (14.5) and 1.3 (0.8), respectively. The QFe values are typical (see the distribution in Figure 2), while the O7O6 value is somewhat smaller (Figure 3).

The solar source of the CME associated with the 1999 September 22 non-MC event is identified by an eruptive filament followed by a post-eruption arcade observed in microwave, soft X-ray and EUV. Figure 16 shows the U-shaped filament at 02:36 UT, which erupts resulting in a two-ribbon flare and post-eruption arcade (PEA) all imaged by the Nobeyama radioheliograph at 17 GHz. The peak brightness temperature (*Tb*) of the PEA in microwaves (17 GHz) is $3.87 \times 10^4$ K. The average brightness temperature of the arcade is $1.53 \times 10^4$ K. The radio emission from the arcade is optically thin, so the kinetic temperature ($T$) of the arcade plasma is given by $T = Tb/\tau$, where $\tau$ is the free-free optical depth of the arcade given by $\tau = 0.2n^2L/f^2T^{3/2}$, where $f$ is the observing frequency (17 GHz), $n$ is the electron density of the arcade plasma, and L is the line-of-sight thickness of the arcade. We need $\tau \leq 0.004$ so that the observed average *Tb* translates into an average kinetic temperature $T \geq 3.5$ MK needed to produce the observed charge states (Bame *et al.*, 1979). Taking the arcade height as its observed width ($L = 9.8 \times 10^9$ cm), one can readily get the required optical depth for an electron density of $(2.-2.2) \times 10^9$ $cm^{-3}$. Such densities have been derived from simultaneous soft X-ray imaging observations in other post-eruption arcades (Hanaoka *et al.*, 1994). Thus, the temperature in the PEA is adequate to produce the observed charge state enhancements in the EP event. We expect a similar situation for most of the EP events and hence conclude that even in such events with poor flare signatures, high charge states can be produced.

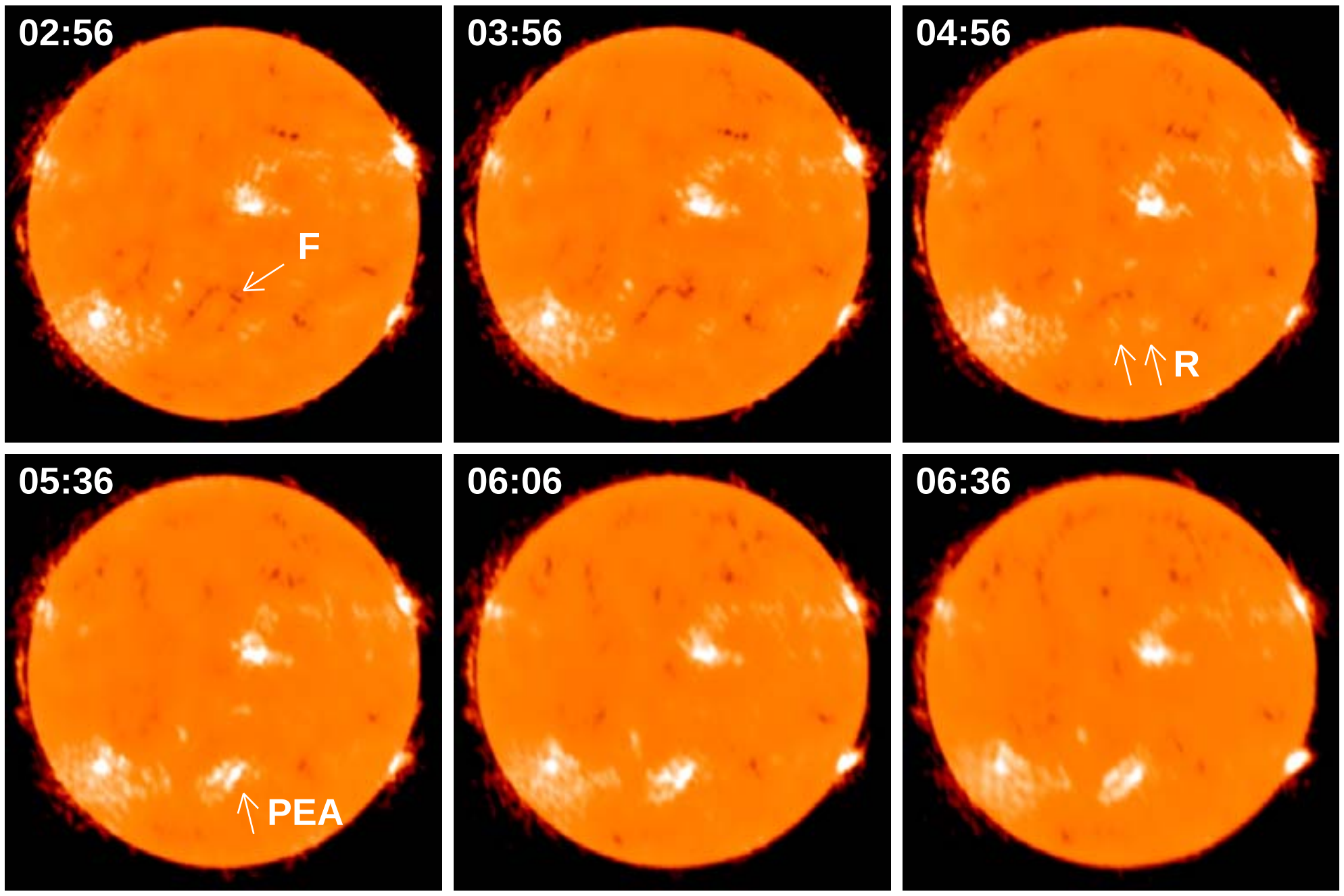

Figure 16. A series of 17 GHz microwave images obtained by the Nobeyama radioheliograph on 1999 September 19 showing the filament (F), its disappearance resulting in a two-ribbon flare (R), and the formation of the post eruption arcade (PEA). The PEA was also observed by Yohkoh/SXT and SOHO/EIT beyond the 06:36 UT (not shown).

We also note that two of the EP events without charge state enhancement are also EJ- events. i.e., we were not able to fit a flux rope event with boundary adjustments. The third EJ- event is the one on 2002 May 30 associated with a C3.7 flare and a filament eruption in the NE quadrant. The filament in the pre-eruption stage (F), the post-eruption arcade (PEA) overlying the filament location, the associated white-light CME, and the GOES soft X-ray light curve are all shown in Figure 17. Note that the white-light CME was clearly surrounded by a shock, but the whole structure is mostly heading to the northwest. In particular, there is only a small section of the CME that crosses the ecliptic, suggesting that the ACE spacecraft measuring the charge states might have passed through only the edge of the ICME. This might be the reason that the observed ICME could not be fit to a flux rope.

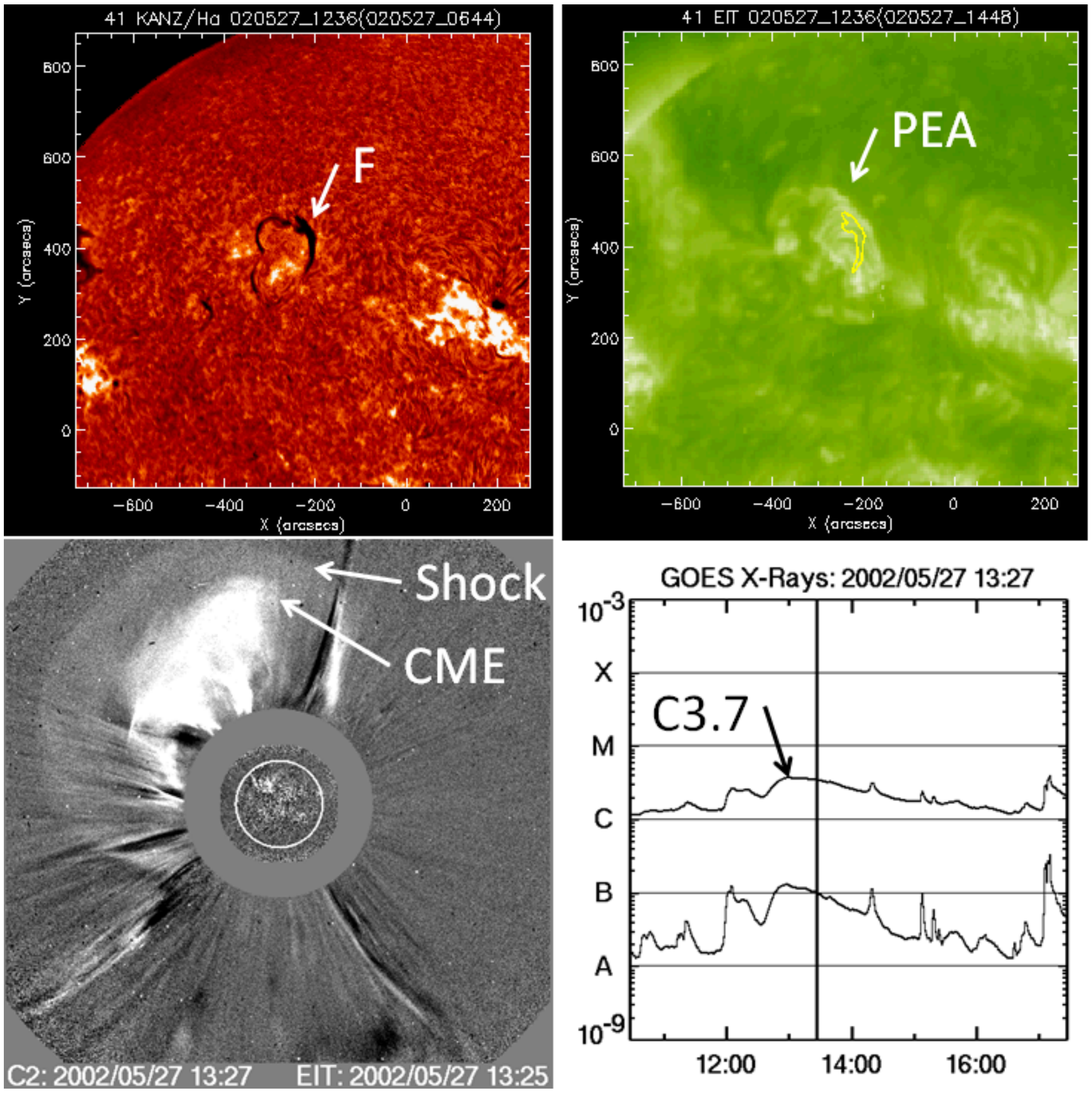

Figure 17. The solar source of the 2002 May 30 non-MC event (one of the three events for which flux rope fitting did not succeed) as a filament (F) eruption event accompanied by a wide shock-driving CME and a weak post-eruption arcade (PEA) responsible for the weak (C3.7) GOES soft X-ray flare on 2002 May 27.

### 3.6 Is there charge state enhancement in the shock sheath?

In a preliminary study, Gopalswamy (2006c) did not find any QFe enhancement in MC sheaths. They found the average QFe in sheaths is ~11.3, which is same as the value in slow solar wind reported by Lepri *et al.* (2001). Figure 18 shows the distribution of QFe in the sheath regions for MC and non-MC events and for the combined set. It is clear that the peak and average QFe in the sheath are enhanced with respect to the threshold values. The enhancement is more prominent in sheaths of MCs than in non-MC sheaths.

**Table 3.** Charge state enhancement in sheaths

| Charge State in Sheath | Event Numbers (Table1) | Fraction | Remark |
|---|---|---|---|
| (i) No enhancement | 4,7, 8, 9, 10, 14, 15, 16, 17, 20, 25, 27, 29, 31, 38, 39, 40, 41, 42, 45, 47, 50, 51, 53, 56, 57, 59 | 27/51 or 53% | |
| (ii) Marginal cases: O7O6 - No, <QFe> - Yes | 5, 54 | 2/51 or 4% | Only QFe enhancement |
| (iii) Marginal cases: O7O6 – Yes, <QFe> - No | 30, 34 | 2/51 or 4% | Only O7O6 enhancement |
| (iv) Enhancement before plasmag starting boundary | 13, 19, 21, 23, 24, 26, 28, 32, 33, 35, 36, 44 | 12/51 or 23% | Charge state signatures precedes plasmag signature in all cases |
| (v) Enhancement due to preceding ICME | 18, 43, 46, 48, 49, 58 | 6/51 or 12% | Plasmag signatures indicate preceding ICME |
| (vi) Other enhancements | 37, 52 | 2/51 or 4% | #52 - marginal enhancement |

In order to examine the charge state enhancements in sheaths, we have listed the events numbers that do and do not show charge state enhancement in sheaths in Table 3. The first three events in Table 1 do not have charge state data, so the remaining 51 are used. First of all we note that more than half of the events (27 out of 51 or 53%) do not have any charge state enhancement in the sheaths. These events are noted as category (i) events in Table 3. Among the remaining 24 events, four (or 8%) were marginal in that only one of QFe and O7O6 showed enhancement in the sheath, that too with just one or two data points above the threshold values (categories ii and iii). Twelve events had charge state enhancements in the tail end of the sheaths. Comparison of the plasmag and charge state signatures revealed that these enhancements can be attributed to the ambiguity in identifying the starting boundary of the ICME based on plasmag signatures. In fact, all these cases, the plasma beta coincided with the onset of charge state enhancement, although there are some short-term fluctuations in the beta value. These events are noted as category (iv) events in Table 3 and add up to 23% of the 51 events. The event shown in Figure 15 is a good example of this type of event. In another six events (marked as category (v) in Table 3), there was definitely preceding ICME material into which the shock is propagating and hence the charge state enhancement can be attributed to the preceding ICME as in Figure 1. Only for two events, marked as category (vi) events in Table 3 that one can say there is charge state enhancement in the sheath. In the case of event #52 (2005 February 15), there were only two consecutive O7O6 data points and a single QFe data point above the respective thresholds. Thus the enhancement is marginal and could be due to fluctuation. In the case of event #37 (2002 April 17), there were two intervals of charge state enhancements, one close to the plasmag starting boundary and the other in the middle of the sheath. The enhancement near the plasmag boundary is similar to that in category (iv) events. However, the enhancement in the middle is during the interval of high beta. Thus, there is only one event among the 51 that can be said to have a charge state enhancement in the sheath. This event needs to be further investigated.

Since the sheath is not connected to the flare site, it is unlikely that the flare plasma enters into the sheath region. Is it possible that the temperature jump across the shock is high enough to enhance the charge state when the shock is very close to the Sun? Comparing the events with no charge state enhancement in Table 3 with their association with type II bursts, we find that more than half of them (15 out of 27) have type II burst association. This means the CMEs were driving strong shocks near the Sun, but there was no charge state enhancement in the sheath. Similarly, there are other events (##23, 44, and 48) that have no type II burst near the Sun (weak shocks) yet they had charge state enhancement. These observations support our conclusion that the temperature jump at the shock may not be related to charge state enhancements observed in the interplanetary medium.

Direct comparison between shock formation observed in EUV images (Gopalswamy *et al.* 2012) and the frequency of the associated metric type II burst suggests that the shock formation can occur at a heliocentric distance as short as 1.2 Rs. The density jump across the shock has been estimated to be only by a factor of ~1.5. If the temperature of the upstream quiet corona is ~1.5 MK, the downstream temperature due to shock heating is expected to be too low to cause the charge state enhancement. Besides, the density in the shock downstream is also expected to be much smaller than in the flare site, which also works against this possibility. However, it must be pointed out that numerical simulation results are not conclusive and give conflicting charge state charge state enhancements in sheaths with respect to the driving CME and the core (see Lynch *et al.* 2011).

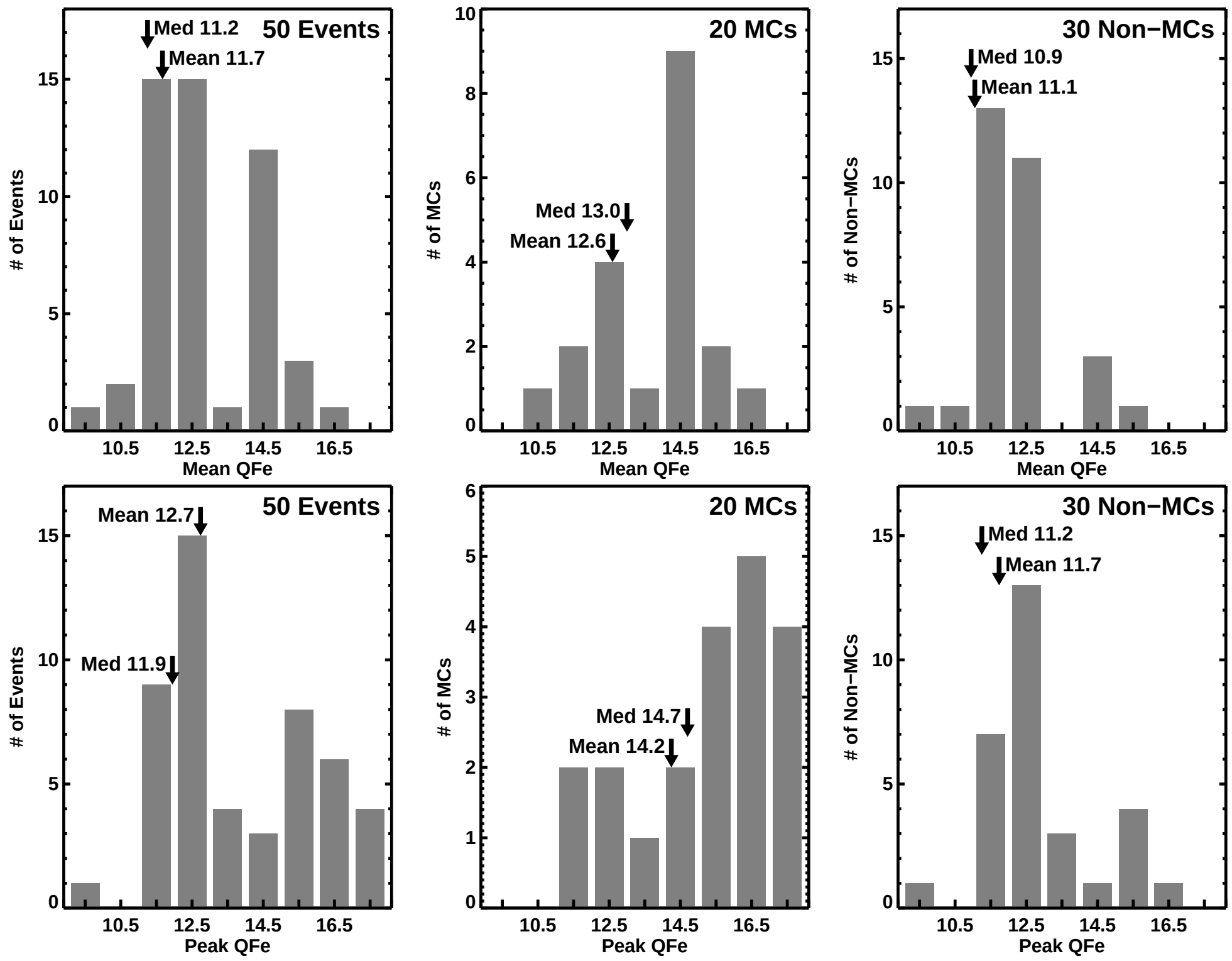

Figure 18. Distribution of average (top) and peak (bottom) QFe values in the sheaths of MC and non-MC ICMEs.

## 4. Discussion

The primary finding of this paper is that the Fe and O charge state measures found inside ICMEs is closely related to the flares that accompany the CMEs. The high temperature resulting from flare heating is responsible for the production of high charge states in the flare plasma, which is injected into the CME flux rope and carried into the IP medium. Charge state enhancement events are excellent examples in which flares and CMEs act in tandem to produce the observed charge state at 1 AU. Without CMEs, the ions cannot get into the IP medium as the charge state data presented here and elsewhere and indicated by models (see *e.g.*, Rakowski, Laming and Lepri, 2007). Two types of magnetic structures are created during an eruptive process: an arcade anchored to the Sun and a flux rope ejected into the heliosphere. This standard model of an eruption elucidated by many authors requires the formation of the two structures, except in confined flares in which all the energy goes into plasma heating and none goes into mass motion (Gopalswamy *et al.*, 2009c). For example, temperatures exceeding 30 MK may be produced in confined flares, but these flares are not accompanied by CMEs (Schmahl *et al.* 1990; Gopalswamy *et al.*, 1995; 2009c). We did not find any significant difference in the flare and CME properties of eruptions associated with MC and non-MC ICMEs. Therefore, there is no obvious reason to expect difference in the topology of the CME structure in the IP medium. The charge state distributions indicate that the charge-state signatures are more prominent in MCs than in non-MC ICMEs. The lower charge state ratio observed in non-MC CMEs can be attributed to the non-radial propagation of the associated CMEs near the Sun, resulting in a less favorable observing geometry. The observing spacecraft does not pass through the axis of the flux rope and thus encounters less of the flare plasma that entered into the flux rope. Such a suggestion was made in Gopalswamy (2006a), which is supported by the charge state analysis presented in this paper.

Many studies have revealed that the high ionization states observed in the IP medium are indicative of a hot source region at the Sun (Bame *et al.*, 1979; Henke *et al.*, 2001; Lepri *et al.*, 2001; Reinard *et al.*, 2001; Reinard 2005; 2008). Apart from the interior of the Sun where thermonuclear reactions occur, one can find temperatures of several to tens of MK only in solar flares. Our analysis finds that the temperature attained in the flaring region ranges from a few MK to 25 MK for both MC and non-MC cases, thus identifying the hot source region on the Sun. The connection between flares and CMEs is that the reconnection produces a flux rope structure (see *e.g.*, Qiu *et al.*, 2007) and the process also injects hot plasma into the flux rope (Lin, Raymond and van Ballegooijen, 2004). The propagation characteristics of the flux rope into the IP medium and how the observing spacecraft passes through the flux rope seem to decide the appearance of the flux rope as an MC or non-MC.

There is considerable observational support that both MCs and non-MCs have a flux rope structure and that the flux ropes associated with non-MCs propagate non-radially. The observational support can be found in the accompanying papers that show that (i) white-light CMEs associated with both MCs and non-MC ICMEs can be fit to flux ropes near the Sun (Xie *et al.* 2012), (ii) propagation direction obtained from the flux rope fit and the CME direction parameter suggest that the CMEs associated with non-MC ICMEs seem to propagate non-radially (Xie *et al.* 2012; Kim *et al.* 2012), (iii) coronal-hole deflection of CMEs is one of the major causes for the non-radial motion of CMEs, and (iv) a flux rope can be fit to even non-MC ICMEs either by slightly modifying the ICME boundaries derived from plasmag signatures or using a torus-type flux rope instead of the conventional cylindrical flux ropes (Marubashi *et al.* 2012). Thus, all evidence points to the conclusion that almost all of the ICMEs reaching far into the IP medium seem to contain a flux rope structure.

The results of this study do not support the idea that some ICMEs may be inherently non-flux ropes, as suggested by Gosling (1990). When active regions slowly expand in to the IP medium, one does not expect flares or mass motions faster than the slow solar wind. In fact, Uchida *et al.* (1992) ruled out that the active region loop expansion involves reconnection. These authors also found that the speed of the expanding loops near the Sun is typically tens of km $s^{-1}$. We saw that almost all the ICME events (MC or non-MC) have charge state enhancements and are associated with flares and fast CMEs. Thus we can rule out active region expansion as a mechanism for non-MC ICMEs (Gosling, 1990). Whether active region expansion leads to any ICMEs is an open question. Antiochos, DeVore, and Klimchuk (1999) speculated that CMEs associated with polar crown filaments may not be CMEs, but loop expansions. However, even CMEs associated with polar crown filaments have post-eruption arcades, similar to the EP events discussed in this paper. Thus we confirm that none of the solar sources of the non-MC ICMEs are active region expansions.

Our study confirms the earlier suggestion by Reinard (2008) that the peripheries of ICMEs may contain weaker charge state signatures. In addition, we think the patchiness of the charge state enhancement within the ICME might contribute to the weaker charge-state signals observed in non-MC ICMEs. The patchy reconnection at the flare site might have contributed to such a situation inside ICMEs. When combined with the fact that the observing spacecraft does not pass through the central axes in the case of non-MC ICMEs, one might expect lower charge state enhancement. The QFe enhancement seems to be more robust that the O7O6 enhancement, probably due to the higher ionization potential of O6+ ions (see also Henke *et al.*, 2001). We find much larger fraction of events with enhanced charge states than in earlier works (Henke *et al.* 2001; Aguilar-Rodriguez, Blanco-Cano, and Gopalswamy 2006) because we have selected events originating from the solar disk center, which seems to be the preferred location for high charge state events (Reinard, 2008). We also find significant overlap between MC and non-MC events in the charge state *vs*. flare properties scatter plots.

In terms of the solar sources, there is one clear difference between the MC and non-MC events: there are far more eruptive prominence and dimming events in the non-MC population (7 *vs*. 1). It is not clear if this is significant because even in these EP events, there are clear flare structures in the form of post-eruption arcades. The temperature attained in these events are also high enough to produce the observed charge states, as illustrated using a case study of the 1999 September 22 non-MC ICME and its solar source.

**5. Summary and Conclusions**

We investigated a set of 54 ICMEs whose solar sources are very close to the disk center (within $\pm 15^{o}$ from the central meridian). The motivation behind this longitude criterion is that CMEs originating from such locations are expected to reach Earth directly and produce MC signatures. More than half of these ICMEs were non-MC events, thus questioning the geometrical hypothesis. We compared the charge state properties at 1 AU between the MC and non-MC events and the corresponding flare and CME properties at the Sun. Our analyses suggest that the MC and non-MC ICMEs have more or less the same eruption characteristics at the Sun. Both types have significant enhancement in charge states. These observations suggest that both MC and non-MC ICMEs are likely to have a flux-rope structure and the observational geometry may be responsible for the appearance of non-MC structures at 1 AU. Specific conclusions of the paper are listed below.

(i) Both MC and non-MC ICMEs are associated with major solar flares, although there are even A and B class flares involved in some cases. The median flare class for non-MC events is slightly smaller than that of the MC events.

(ii) The flare temperatures derived from GOES soft X-ray data are in the range 5 – 25 MK for both MC and non-MC events. Even in the case of eruptive prominence events in which the flare temperature could not be derived from GOES data, there is radio evidence of flare temperature high enough to produce the observed charge states.

(iii) The CME properties are similar between MC and non-MC events in terms of their sky-plane speed, width, and acceleration. The CMEs are more energetic than ordinary CMEs. The fraction of halo CMEs in the two populations is very high, exceeding 70%.

(iv) There is good correlation between Fe and O charge state enhancements in ICMEs and the flare properties such as soft X-ray peak flux and flare temperature. The correlation with CME speed is moderate for Fe charge states, but poor for O charge states. CMEs are not directly involved in the production of high charge states, so the observed correlation simply reflects the correlation between CME kinetic energy and soft X-ray peak flux known before (see, *e.g.*, Yashiro and Gopalswamy 2009).

(v) There is significant difference in the boundaries derived from the solar wind plasma and magnetic signatures and from the charge signatures: the charge state signatures systematically start before the starting ICME boundary This may be responsible for the enhanced charge states observed in many ICME sheaths. Charge state enhancements in shock sheaths are also found when the shock moves through a preceding ICME. There is only one clear case in our sample in which true charge state enhancement was found in the sheath and needs further investigation.

(vi) The durations of charge state enhancement above the Fe and O thresholds is considerably smaller than the ICME duration derived from the solar wind plasma and magnetic signatures. This suggests that the charge state enhancement within the ICMEs is patchy.

(vii) Combined with the results of the accompanying papers, we find that CMEs associated with non-MC ICMEs are prone to deflection by coronal holes resulting in non-radial propagation, which might have contributed to the observation of non-MCs at 1 AU. The patchiness of enhanced charge state in CMEs also lowers the probability of observing the flux rope structure at 1 AU.

(viii) We conclude that the production mechanism for high charge states and the flux rope structure are the same for MC and non-MC ICMEs. However, the observing geometry is different, resulting from propagation differences.

(ix) We do not find any evidence for active region expansion resulting in ICMEs lacking a flux rope structure.

**Acknowledgements**

We thank the ACE, Wind and SOHO teams for providing the data on line. SOHO is a project of international collaboration between ESA and NASA.

**Table 1. List of ICMEs originating from the disk center with the solar source and 1 AU charge state information**

| Event #[a] | Shock | | ICME | | | | | CME | | | | | Solar Source | | | Type II? | Charge States | | | | | | | | | | | |
|---|---|---|---|---|---|---|---|---|---|---|---|---|---|---|---|---|---|---|---|---|---|---|---|---|---|---|---|---|
| | | | | Start | | End | | Onset | | | | | | | | | QFe | | | | | | $O^{+7}/O^{+6}$ | | | | | |
| | Date | Time UT | Type[b] | Date mm/dd | Time UT | Date mm/dd | Time UT | Date | Time UT | Width deg | Speed km/s | Acc. $m/s^2$ | Onset UT | Loc | Flare Imp.[c] | | Peak | Ave | Dur1[d] | Dur2[e] | Fr1[f] | Fr2[g] | Peak | Ave | Dur1[d] | Dur2[e] | Fr1[f] | Fr2[g] |
| 1 | 1997/01/10 | 00:52 | MC | 01/10 | 05:18 | 01/11 | 02:18 | 01/06 | 15:10 | 360 | 136 | 4.1 | 14:54 | S18E06 | A1.1 | Yes | ---- | ---- | ---- | ---- | ---- | ---- | ---- | ---- | ---- | ---- | ---- | ---- |
| 2 | 1997/05/15 | 01:15 | MC | 05/15 | 09:06 | 05/16 | 01:06 | 05/12 | 05:30 | 360 | 464 | -15.0 | 04:42 | N21W08 | C1.3 | Yes | ---- | ---- | ---- | ---- | ---- | ---- | ---- | ---- | ---- | ---- | ---- | ---- |
| 3 | 1997/12/10 | 04:30 | EJ+ | 12/11 | 03:45 | 12/11 | 09:00 | 12/06 | 10:27 | 223 | 397 | 9.0 | 10:00 | N45W10 | EP | Yes | ---- | ---- | ---- | ---- | ---- | ---- | ---- | ---- | ---- | ---- | ---- | ---- |
| 4 | 1998/05/03 | 17:00 | EJ? | 05/03 | 19:00 | 05/04 | 00:00 | 05/01 | 23:40 | 360 | 585 | 8.0 | 22:36 | S18W05 | M1.2 | No | 10.6 | 9.2 | 14.0 | ---- | 2.80 | ---- | 0.6 | 0.4 | ---- | ---- | ---- | ---- |
| 5 | 1998/05/04 | 02:00 | EJ+ | 05/04 | 10:00 | 05/05 | 01:15 | 05/02 | 14:06 | 360 | 938 | -28.8 | 13:31 | S15W15 | X1.1 | Yes | 15.6 | 15.5 | 4.0 | 4.0 | 0.26 | 0.26 | 0.7 | 0.6 | 15.0 | 9.0 | 0.98 | 0.59 |
| 6 | 1998/06/25 | 16:10 | EJ+ | 06/26 | 02:00 | 06/26 | 19:00 | 06/22 | 07:34 | 119 | 278 | 6.7 | 05:56 | S28W35 | EP | No | 12.3 | 11.6 | ---- | ---- | ---- | ---- | 0.6 | 0.4 | ---- | ---- | ---- | ---- |
| 7 | 1998/11/07 | 08:00 | EJ+ | 11/07 | 22:00 | 11/08 | 02:00 | 11/04 | 07:54 | 360 | 523 | 19.6 | 07:13 | N17W01 | C1.6 | No | 11.8 | 11.0 | 37.5 | ---- | 9.37 | ---- | 0.7 | 0.5 | 1.0 | 1.0 | 0.25 | 0.25 |
| 8 | 1998/11/13 | 01:40 | EJ+ | 11/13 | 04:30 | 11/14 | 10:15 | 11/09 | 18:18 | 190 | 325 | 2.6 | 17:03 | N15W05 | C2.5 | Yes | 13.4 | 10.9 | 2.0 | 2.0 | 0.07 | 0.07 | 0.5 | 0.3 | ---- | ---- | ---- | ---- |
| 9 | 1999/04/16 | 11:10 | MC | 04/16 | 20:18 | 04/17 | 21:18 | 04/13 | 03:30 | 261 | 291 | 0.2 | 01:45 | N16E00 | B4.3 | No | 14.9 | 13.3 | 27.9 | 22.0 | 1.12 | 0.88 | 1.0 | 0.6 | 16.0 | 11.0 | 0.64 | 0.44 |
| 10 | 1999/06/26 | 19:25 | EJ+ | 06/27 | 21:30 | 06/28 | 01:00 | 06/24 | 13:31 | 360 | 975 | 32.4 | 12:04 | N29W13 | C4.1 | Yes | 12.2 | 11.9 | 8.0 | 2.0 | 2.29 | 0.57 | 0.4 | 0.3 | ---- | ---- | ---- | ---- |
| 11 | 1999/07/02 | 00:23 | EJ+ | 07/02 | 06:00 | 07/02 | 07:30 | 06/29 | 19:54 | 360 | 560 | -8.9 | 19:07 | S14E01 | M1.6 | Yes | ---- | ---- | ---- | ---- | ---- | ---- | 0.3 | 0.3 | ---- | ---- | ---- | ---- |
| 12 | 1999/08/08 | 17:44 | MC | 08/09 | 10:48 | 08/10 | 15:48 | 08/02 | 07:26 | 189 | 286 | -0.5 | 06:12 | N13E24 | EP | ? | 11.1 | 10.3 | ---- | ---- | ---- | ---- | 0.9 | 0.6 | ---- | ---- | ---- | ---- |
| 13 | 1999/09/22 | 12:00 | EJ+ | 09/22 | 21:00 | 09/24 | 02:00 | 09/20 | 06:06 | 360 | 604 | -14.5 | 03:58 | S20W05 | EP | No | 15.6 | 14.5 | 44.0 | 30.0 | 1.52 | 1.03 | 1.3 | 0.8 | 28.0 | 21.0 | 0.97 | 0.72 |
| 14 | 1999/10/21 | 02:13 | EJ+ | 10/21 | 18:30 | 10/22 | 05:50 | 10/18 | 00:06 | 240 | 144 | 3.5 | 23:22 | S30E15 | C1.2 | No | 14.4 | 13.0 | 14.0 | 8.0 | 1.24 | 0.71 | 0.8 | 0.5 | 1.0 | 3.0 | 0.09 | 0.26 |
| 15 | 2000/01/22 | 00:23 | EJ+ | 01/22 | 18:00 | 01/23 | 02:00 | 01/18 | 17:54 | 360 | 739 | -7.1 | 17:07 | S19E11 | M3.9 | Yes | 10.8 | 10.5 | 4.0 | ---- | 0.50 | 0.00 | 0.6 | 0.3 | 7.0 | ---- | 0.88 | ---- |
| 16 | 2000/02/20 | 21:00 | MC | 02/21 | 09:48 | 02/22 | 13:18 | 02/17 | 21:30 | 360 | 728 | -22.9 | 20:17 | S29E07 | M1.3 | Yes | 16.8 | 14.6 | 41.9 | 22.0 | 1.52 | 0.80 | 1.3 | 0.6 | 17.0 | 13.0 | 0.62 | 0.47 |
| 17 | 2000/07/10 | 06:00 | EJ+ | 07/11 | 01:30 | 07/11 | 11:22 | 07/07 | 10:26 | 360 | 453 | 10.8 | 06:24 | N04E00 | EP | No | 14.2 | 12.9 | 14.1 | 6.0 | 1.43 | 0.61 | 3.1 | 1.6 | 18.0 | 9.0 | 1.82 | 0.91 |
| 18 | 2000/07/11 | 11:22 | EJ+ | 07/11 | 22:48 | 07/13 | 02:25 | 07/08 | 23:50 | 161 | 483 | -7.2 | 22:58 | N18W12 | C4.0 | Yes | 16.4 | 14.9 | 26.1 | 24.0 | 0.94 | 0.87 | 2.0 | 1.0 | 22.0 | 21.0 | 0.80 | 0.76 |
| 19 | 2000/07/15 | 14:18 | MC | 07/15 | 21:06 | 07/16 | 09:54 | 07/14 | 10:54 | 360 | 1674 | -96.1 | 10:03 | N22W07 | X5.7 | Yes | 17.7 | 16.9 | 15.8 | 12.0 | 1.23 | 0.94 | 2.0 | 1.2 | 17.0 | 10.0 | 1.33 | 0.78 |
| 20 | 2000/07/26 | 18:58 | EJ+ | 07/27 | 08:28 | 07/27 | 19:35 | 07/23 | 05:30 | 181 | 631 | -20.4 | 04:11 | S13W05 | EP | No | ---- | ---- | ---- | ---- | ---- | ---- | 1.1 | 0.8 | 9.0 | 10.0 | 0.81 | 0.90 |
| 21 | 2000/07/28 | 06:39 | MC | 07/28 | 21:06 | 07/29 | 10:06 | 07/25 | 03:30 | 360 | 528 | -5.8 | 02:43 | N06W08 | M8.0 | Yes | 15.8 | 12.8 | 11.9 | 8.0 | 0.92 | 0.62 | 4.1 | 1.0 | 10.0 | 6.0 | 0.77 | 0.46 |
| 22 | 2000/08/10 | 05:10 | EJ+ | 08/10 | 19:00 | 08/11 | 12:00 | 08/06 | 23:06 | 40 | 597 | -7.0 | 22:36 | S24W15 | ? | No | 11.1 | 10.7 | ---- | ---- | ---- | ---- | 0.9 | 0.8 | ---- | ---- | ---- | ---- |
| 23 | 2000/08/11 | 18:51 | MC | 08/12 | 06:06 | 08/13 | 05:06 | 08/09 | 16:30 | 360 | 702 | 2.8 | 15:19 | N20E12 | EP | No | 15.4 | 13.6 | 38.2 | 20.0 | 1.66 | 0.87 | 2.5 | 0.9 | 37.0 | 13.0 | 1.61 | 0.57 |
| 24 | 2000/09/17 | 17:00 | MC | 09/18 | 01:54 | 09/18 | 15:06 | 09/16 | 05:18 | 360 | 1215 | -12.3 | 04:06 | N14W07 | M5.9 | Yes | 17.3 | 15.9 | 47.9 | 14.0 | 3.63 | 1.06 | 5.3 | 2.0 | 22.0 | 12.0 | 1.67 | 0.91 |
| 25 | 2000/10/05 | 03:23 | EJ+ | 10/05 | 13:13 | 10/07 | 13:00 | 10/02 | 03:50 | 360 | 525 | -4.9 | 02:48 | S09E07 | C4.1 | No | 12.7 | 11.7 | 37.1 | 12.0 | 0.78 | 0.25 | 0.6 | 0.4 | ---- | 2.0 | ---- | 0.04 |
| 26 | 2000/10/12 | 22:36 | MC | 10/13 | 18:24 | 10/14 | 16:54 | 10/09 | 23:50 | 360 | 527 | -24.2 | 23:19 | N01W14 | C6.7 | Yes | 16.1 | 14.4 | 19.8 | 20.0 | 0.88 | 0.89 | 1.3 | 0.8 | 15.0 | 18.0 | 0.67 | 0.80 |
| 27 | 2000/11/06 | 09:20 | MC | 11/06 | 23:06 | 11/07 | 18:06 | 11/03 | 18:26 | 360 | 291 | 16.4 | 18:35 | N02W02 | C3.2 | Yes | 14.8 | 12.5 | 44.0 | 12.0 | 2.32 | 0.63 | 2.1 | 0.9 | 10.0 | 13.0 | 0.53 | 0.68 |
| 28 | 2000/11/26 | 05:30 | EJ+ | 11/27 | 05:00 | 11/28 | 04:00 | 11/24 | 05:30 | 360 | 1289 | 2.1 | 04:55 | N20W05 | X2.0 | Yes | 15.6 | 13.7 | 60.0 | 24.0 | 2.61 | 1.04 | 1.4 | 0.8 | 17.0 | 18.0 | 0.74 | 0.78 |
| 29 | 2001/03/03 | 11:30 | EJ+ | 03/04 | 04:00 | 03/05 | 01:30 | 02/28 | 14:50 | 232 | 313 | 1.9 | 13:22 | S17W05 | B4.2 | No | 14.5 | 12.1 | 8.0 | 8.0 | 0.37 | 0.37 | 1.2 | 0.8 | 6.0 | 17.0 | 0.28 | 0.79 |
| 30 | 2001/03/22 | 14:00 | EJ+ | 03/22 | 22:30 | 03/23 | 04:00 | 03/19 | 05:26 | 360 | 389 | -2.4 | 04:12 | S20W00 | PEA | No | 9.8 | 9.7 | ---- | ---- | ---- | ---- | 0.4 | 0.3 | ---- | ---- | ---- | ---- |
| 31 | 2001/04/11 | 14:12 | EJ+ | 04/11 | 22:30 | 04/12 | 03:00 | 04/09 | 15:54 | 360 | 1192 | 1.3 | 15:20 | S21W04 | M7.9 | Yes | 15.3 | 15.1 | 10.0 | 4.0 | 2.23 | 0.89 | 4.8 | 1.9 | 7.0 | 5.0 | 1.56 | 1.11 |
| 32 | 2001/04/11 | 16:19 | MC | 04/12 | 07:54 | 04/12 | 17:54 | 04/10 | 05:30 | 360 | 2411 | 211.6 | 05:06 | S23W09 | X2.3 | Yes | 16.2 | 14.6 | 22.0 | 8.0 | 2.20 | 0.80 | 1.3 | 1.0 | 24.0 | 9.0 | 2.40 | 0.90 |
| 33 | 2001/04/28 | 05:02 | MC | 04/29 | 01:54 | 04/29 | 12:54 | 04/26 | 12:30 | 360 | 1006 | 21.1 | 11:26 | N20W05 | M1.5 | Yes | 16.7 | 15.9 | 48.0 | 12.0 | 4.36 | 1.09 | 2.0 | 1.5 | 47.0 | 9.0 | 4.27 | 0.82 |
| 34 | 2001/08/12 | 11:10 | EJ- | 08/13 | 07:00 | 08/13 | 10:00 | 08/09 | 10:30 | 175 | 479 | 4.4 | 08:00 | N11W14 | PEA | Yes | 12.6 | 11.6 | 2.0 | 2.0 | 0.67 | 0.67 | 0.3 | 0.2 | ---- | ---- | ---- | ---- |
| 35 | 2001/10/11 | 16:50 | EJ+ | 10/12 | 03:30 | 10/12 | 08:30 | 10/09 | 11:30 | 360 | 973 | -41.5 | 10:46 | S28E08 | M1.4 | Yes | 15.1 | 14.0 | 6.0 | 6.0 | 1.20 | 1.20 | 1.7 | 1.0 | 2.0 | 5.0 | 0.40 | 1.00 |
| 36 | 2002/03/18 | 13:13 | MC | 03/19 | 22:54 | 03/20 | 15:24 | 03/15 | 23:06 | 360 | 957 | -17.4 | 22:09 | S08W03 | M2.2 | Yes | 13.7 | 13.7 | ---- | 2.0 | ---- | 0.12 | 1.5 | 1.0 | 18.0 | 12.0 | 1.09 | 0.73 |
| 37 | 2002/04/17 | 11:01 | MC | 04/18 | 04:18 | 04/19 | 02:18 | 04/15 | 03:50 | 360 | 720 | 2.1 | 03:05 | S15W01 | M1.2 | Yes | 14.9 | 14.6 | 22.0 | 4.0 | 1.00 | 0.18 | 4.1 | 2.2 | 16.0 | 11.0 | 0.73 | 0.50 |
| 38 | 2002/05/11 | 10:30 | EJ+ | 05/11 | 13:00 | 05/11 | 14:00 | 05/08 | 13:50 | 360 | 614 | 78.9 | 12:58 | S12W07 | C4.2 | No | 11.4 | 11.4 | 6.0 | ---- | 6.01 | ---- | 0.3 | 0.3 | 4.0 | ---- | 4.00 | ---- |
| 39 | 2002/05/18 | 19:51 | MC | 05/19 | 03:54 | 05/19 | 23:24 | 05/16 | 00:50 | 360 | 600 | -6.6 | 00:11 | S23E15 | C4.5 | Yes | 10.7 | 10.1 | ---- | ---- | ---- | ---- | 0.2 | 0.1 | ---- | ---- | ---- | ---- |
| 40 | 2002/05/20 | 03:40 | EJ- | 05/20 | 11:00 | 05/20 | 22:00 | 05/17 | 01:27 | 45 | 461 | 5.5 | 00:23 | S20E14 | EP | No | 11.6 | 11.1 | ---- | ---- | ---- | ---- | 0.4 | 0.2 | ---- | ---- | ---- | ---- |
| 41 | 2002/05/30 | 02:15 | EJ- | 05/30 | 07:09 | 05/31 | 11:20 | 05/27 | 13:27 | 161 | 1106 | 3.8 | 12:36 | N22E15 | C3.7 | No | 12.1 | 11.0 | 2.0 | 2.0 | 0.07 | 0.07 | 0.7 | 0.5 | ---- | 3.0 | ---- | 0.11 |
| 42 | 2002/07/17 | 15:50 | EJ+ | 07/18 | 12:00 | 07/19 | 08:10 | 07/15 | 21:30 | 188 | 1300 | -7.3 | 21:03 | N19W01 | M1.8 | Yes | ---- | ---- | ---- | ---- | ---- | ---- | 1.1 | 0.6 | 15.0 | 9.0 | 0.74 | 0.45 |
| 43 | 2002/08/01 | 05:10 | MC | 08/01 | 11:54 | 08/01 | 22:36 | 07/29 | 12:07 | 161 | 222 | 3.3 | 10:27 | S10W10 | M4.7 | Yes | 15.2 | 13.5 | 17.9 | 12.0 | 1.67 | 1.12 | 2.6 | 1.3 | 17.0 | 10.0 | 1.59 | 0.93 |
| 44 | 2003/08/17 | 13:40 | MC | 08/18 | 11:36 | 08/19 | 04:24 | 08/14 | 20:06 | 360 | 378 | 4.4 | 17:12 | S10E02 | C3.8 | No | 14.3 | 12.9 | 37.7 | 14.0 | 2.24 | 0.83 | 2.1 | 1.4 | 37.0 | 17.0 | 2.20 | 1.01 |
| 45 | 2003/10/29 | 06:00 | MC | 10/29 | 08:00 | 10/30 | 04:00 | 10/28 | 11:30 | 360 | 2459 | -105.2 | 11:00 | S16E08 | X17.2 | Yes | 17.0 | 15.3 | 60.1 | 18.0 | 3.01 | 0.90 | 1.2 | 0.6 | 34.0 | 12.0 | 1.70 | 0.60 |
| 46 | 2003/10/30 | 16:20 | MC | 10/31 | 02:00 | 10/31 | 13:00 | 10/29 | 20:54 | 360 | 2029 | -146.5 | 20:37 | S15W02 | X10.0 | Yes | 17.3 | 16.7 | 48.0 | 10.0 | 4.36 | 0.91 | 8.5 | 3.0 | 45.0 | 8.0 | 4.09 | 0.73 |
| 47 | 2004/01/22 | 01:10 | EJ+ | 01/22 | 08:00 | 01/23 | 17:00 | 01/20 | 00:06 | 360 | 965 | 17.2 | 23:46 | S13W09 | C5.5 | No | 14.2 | 11.6 | 13.8 | 12.0 | 0.42 | 0.36 | 3.1 | 0.8 | 21.0 | 16.0 | 0.64 | 0.48 |
| 48 | 2004/07/24 | 05:32 | MC | 07/24 | 12:48 | 07/25 | 13:18 | 07/22 | 08:30 | 132 | 899 | -12.6 | 07:41 | N04E10 | C5.3 | No | 14.5 | 13.5 | 52.0 | 6.0 | 2.12 | 0.24 | 3.2 | 1.5 | 50.0 | 19.0 | 2.04 | 0.78 |
| 49 | 2004/11/09 | 09:05 | MC | 11/09 | 20:54 | 11/10 | 03:24 | 11/06 | 02:06 | 214 | 1111 | 18.8 | 01:40 | N09E05 | M3.6 | Yes | 13.7 | 13.1 | 48.0 | 6.0 | 7.39 | 0.92 | 1.3 | 0.8 | 48.0 | 4.0 | 7.38 | 0.62 |
| 50 | 2004/12/11 | 13:03 | EJ+ | 12/12 | 12:00 | 12/13 | 06:00 | 12/08 | 20:26 | 360 | 611 | -87.2 | 19:34 | N05W03 | C2.5 | Yes | 13.9 | 11.4 | 21.6 | 6.0 | 1.20 | 0.33 | 0.6 | 0.4 | 5.0 | ---- | 0.28 | ---- |
| 51 | 2005/01/16 | 09:27 | EJ+ | 01/16 | 14:00 | 01/17 | 06:30 | 01/15 | 06:30 | 360 | 2049 | -30.7 | 05:54 | N16E04 | M8.6 | Yes | 15.1 | 13.2 | 34.8 | 16.0 | 2.11 | 0.97 | 0.8 | 0.3 | ---- | 1.0 | ---- | 0.06 |
| 52 | 2005/02/17 | 21:59 | EJ+ | 02/18 | 15:00 | 02/19 | 08:15 | 02/13 | 11:06 | 151 | 584 | -13.0 | 10:28 | S11E09 | C2.7 | Yes | 14.6 | 13.5 | 14.0 | 12.0 | 0.81 | 0.70 | 1.2 | 0.6 | 5.0 | 8.0 | 0.29 | .46 |
| 53 | 2005/05/15 | 02:19 | MC | 05/15 | 05:42 | 05/15 | 22:12 | 05/13 | 17:12 | 360 | 1689 | ----- | 16:13 | N12E11 | M8.0 | Yes | 16.2 | 14.7 | 30.0 | 18.0 | 1.82 | 1.70 | 3.7 | 1.6 | 28.0 | 17.0 | 1.70 | 1.03 |
| 54 | 2005/05/20 | 03:34 | MC | 05/20 | 07:18 | 05/21 | 05:18 | 05/17 | 03:26 | 273 | 449 | 18.1 | 02:31 | S15W00 | M1.8 | Yes | 16.8 | 14.7 | 21.8 | 18.0 | 0.99 | 0.82 | 2.1 | 1.0 | 37.0 | 18.0 | 1.68 | 0.82 |
| 55 | 2005/05/29 | 09:15 | EJ+ | 05/29 | 10:15 | 05/29 | 14:45 | 05/26 | 15:06 | 360 | 586 | -1.6 | 13:10 | S11E19 | B7.5 | ? | 11.2 | 11.0 | ---- | ---- | ---- | ---- | 0.7 | 0.4 | ---- | ---- | ---- | ---- |
| 56 | 2005/07/10 | 02:56 | EJ+ | 07/10 | 10:30 | 07/12 | 04:00 | 07/07 | 17:06 | 360 | 683 | -8.7 | 16:07 | N09E03 | M4.9 | Yes | 16.8 | 13.5 | 50.0 | 24.0 | 1.21 | 0.58 | 1.6 | 0.7 | 43.0 | 17.0 | 1.04 | 0.41 |
| 57 | 2005/09/02 | 13:32 | EJ+ | 09/02 | 19:03 | 09/03 | 06:00 | 08/31 | 11:30 | 360 | 825 | 42.9 | 10:26 | N13W13 | C2.0 | Yes | 12.6 | 10.9 | 2.0 | 2.0 | 0.18 | 0.18 | 0.5 | 0.2 | ---- | ---- | ---- | ---- |
| 58 | 2005/09/15 | 08:25 | EJ+ | 09/15 | 14:24 | 09/15 | 18:00 | 09/13 | 20:00 | 360 | 1866 | 11.5 | 19:42 | S09E10 | X1.5 | Yes | 14.1 | 14.1 | 6.0 | 2.0 | 1.67 | 0.56 | 0.5 | 0.4 | ---- | ---- | ---- | ---- |
| 59 | 2006/08/19 | 10:51 | EJ+ | 08/20 | 00:00 | 08/21 | 15:30 | 08/16 | 16:30 | 360 | 888 | 1.9 | 14:37 | S16W08 | C3.6 | Yes | 15.1 | 14.0 | 8.0 | 8.0 | 0.20 | 0.20 | 0.6 | 0.3 | 1.0 | 1.0 | 0.03 | 0.03 |

[a] List of shock-driving ICMEs during the solar cycle 23 (E15° ≤ source longitude ≤ W15°) (Gopalswamy et al. 2010a).
[b] MC = Magnetic cloud; EJ = Ejecta; the suffix + indicates that it was possible fit a flux rope to the ejecta by adjusting the plasmag boundaries; - indicates it was not possible to fit a flux rope
[c] EP = Eruptive prominence; PEA = post-eruption arcade
[d] dur1 = duration of charge state enhancements, without considering the second plasmag boundary
[e] dur2 = duration of charge state enhancements, within the plasmag boundaries
[d] Fr1 = Fractional duration of charge state enhancements with respect to the plasmag ICME duration
[e] Fr2 = Fractional duration of charge state enhancements within the plasmag boundaries, considering only intervals during which the charge states are above the thresholds

#6, #12, #55 Dropped from the analysis because the revised solar source location fell outside the longitude criterion.
#11 Dropped from the analysis because this is a known "driverless" event.
#22 Dropped from the analysis because of the uncertainty in identifying the solar source; multiple candidates exist.